\documentclass[times]{aastex631}
\usepackage{amsmath}
\usepackage{bm}
\usepackage{hyperref}

\begin{document}
\title{X-Ray Radiative Transfer Calculation Based on a Physics-based Model of \\ 
the Sub-parsec-scale Gases around an Active Galactic Nucleus and Its Application to NGC~3783}
\correspondingauthor{Atsushi Tanimoto}
\email{atsushi.tanimoto@sci.kagoshima-u.ac.jp}
\author[0000-0002-0114-5581]{Atsushi Tanimoto}
\affiliation{Graduate School of Science and Engineering, Kagoshima University, Kagoshima 890-0065, Japan}
\author[0000-0002-8779-8486]{Keiichi Wada}
\affiliation{Graduate School of Science and Engineering, Kagoshima University, Kagoshima 890-0065, Japan}
\affiliation{Research Center for Space and Cosmic Evolution, Ehime University, Matsuyama 790-8577, Japan}
\affiliation{Faculty of Science, Hokkaido University, Sapporo 060-0810, Japan}
\author[0000-0003-2670-6936]{Hirokazu Odaka}
\affiliation{Department of Earth and Space Science, Osaka University, Osaka 560-0043, Japan}
\author[0000-0003-0548-1766]{Yuki Kudoh}
\affiliation{Astronomical Institute, Tohoku University, Miyagi 980-8578, Japan}
\author[0000-0003-2535-5513]{Nozomu Kawakatu}
\affiliation{Faculty of Natural Sciences, National Institute of Technology, Kure College, Hiroshima 737-8506, Japan}

\begin{abstract}
Although the X-ray spectra of Seyfert~1 galaxies exhibit absorption lines of He-like iron and H-like iron at blue-shifted velocities of approximately $500 \ \mathrm{km} \ \mathrm{s}^{-1}$, the physical origin of these absorption lines remains uncertain. In this study, we performed X-ray radiative transfer based on the sub-parsec-scale thermally driven outflows. The initial step involved calculating the photoionization equilibrium using the Cloudy code, which is based on three-dimensional radiative hydrodynamic simulations. Subsequently, X-ray radiative transfer was performed using the Monte Carlo simulation for astrophysics and cosmology code. Our findings indicate that when the angle of inclination ranges from $55\degr$ to $65\degr$, the transmitted component of the X-ray spectrum displays absorption lines of He-like and H-like iron, exhibiting a blue shift of approximately $500 \ \mathrm{km} \ \mathrm{s}^{-1}$. The results suggest that the absorption lines are generated by a photoionized gas within $0.005 \ \mathrm{pc}$. Additionally, the results indicate that the scattered component of the X-ray spectrum exhibits emission lines originating from neutral iron fluorescence, He-like iron, and H-like iron. The emission lines are broadened by approximately $7000 \ \mathrm{km} \ \mathrm{s}^{-1}$ due to the Keplerian rotation. Furthermore, the model reproduced the H-like iron and H-like iron absorption lines in NGC~3783 observed by the Chandra High Energy Transmission Grating.
\end{abstract}.
\section{Introduction}\label{Section0101}
Multi-wavelength observations indicate that at least half of active galactic nuclei (AGNs) exhibit outflows, implying that outflows may be a universal phenomenon \citep[e.g.,][]{Laha2021}. AGN outflows play a crucial role in the co-evolution between supermassive black holes (SMBHs) and their host galaxies \citep[e.g.,][]{Kormendy2013}. This is because the AGN outflow has the potential to heat the gas in the host galaxy and stop its star formation activity \citep[e.g.,][]{Fabian2012}. However, the impact of AGN outflows on their host galaxies and the mechanisms that drive AGN outflows remain poorly understood.

One of the most crucial physical quantities for understanding the driving mechanism of the AGN outflow is the Eddington ratio. The reason is that the Eddington ratio represents the radiation pressure of the AGN continuum relative to the gravity of the SMBH. When the Eddington ratio is greater than unity, the radiation pressure from the AGN continuum is greater than the gravity of the SMBH, which naturally drives the AGN outflow. To understand the Eddington ratio of AGN with outflows, \cite{Ganguly2007} examined the distribution of Eddington ratios for approximately $500$ quasars observed by the Sloan Digital Sky Survey. They found that approximately $80\%$ of the objects exhibited Eddington ratios less than unity. This result indicated that it is difficult to drive the outflow only by the radiation pressure of the AGN continuum. Therefore, it is essential to consider additional driving mechanisms such as the ultraviolet line force \citep{Proga2000, Proga2004, Proga2007, Proga2008, Nomura2016, Nomura2017, Nomura2020}, the pressure gradient force by X-ray heating \citep{Balsara1993, Dorodnitsyn2008a, Dorodnitsyn2008b, Dorodnitsyn2009, Kurosawa2009, Sim2012, Done2018, Mizumoto2019, Dannen2020, Ganguly2021, Water2021}, the radiation pressure to the gas containing dust \citep{Wada2012, Wada2016, Kudoh2023}, and the magnetic centrifugal force \citep{Fukumura2010, Fukumura2015, Fukumura2018, Kallman2019}.

One of the most effective methods for investigating the driving mechanism of the AGN outflow is precision X-ray spectroscopy. This is because X-rays are capable of detecting absorption lines resulting from the presence of partially ionized materials. For example, the outflow velocity can be determined by measuring the deviation between the observed energy of the absorption lines and that of a static system. In fact, Chandra High Energy Transmission Grating (HETG; \citealt{Canizares2005}) and the XMM-Newton Reflection Grating Spectrometer (RGS; \citealt{Herder2001}) detected many blue-shifted absorption lines in the X-ray spectra of nearby AGNs with velocities ranging from $100 \ \mathrm{km} \ \mathrm{s}^{-1}$ to $3000 \ \mathrm{km} \ \mathrm{s}^{-1}$ \citep[e.g.,][]{McKernan2007, Laha2014}. By comparing these X-ray spectroscopic observations with the results of hydrodynamic simulations, we can investigate whether or not any mechanism is appropriate for driving the AGN outflow. However, in order to directly compare X-ray spectroscopic observations with hydrodynamic simulations, it is essential to perform X-ray radiative transfer calculations based on simulations. 

\cite{Ogawa2022} compared the X-ray spectroscopic observations of a Seyfert 1 galaxy with a radiation pressure model, called the radiation-driven fountain model \citep{Wada2016}. \cite{Wada2016} showed that the unsteady AGN outflow is driven by the radiation pressure on the dusty gas using a parsec-scale three-dimensional radiative hydrodynamic simulation. Since the cross section of dust to Mie scattering is more than two orders of magnitude larger than that for Thomson scattering \citep{Draine2003}, the radiation pressure on the dusty gas can drive the AGN outflow even in the case of sub-Eddington accretion. \cite{Ogawa2022} calculated the X-ray spectrum based on the parsec-scale radiation-driven fountain model using the Cloudy photoionization code \citep{Ferland2017} and compared their model with the X-ray spectrum of the Seyfert~1 galaxy NGC~4051 observed with the XMM-Newton RGS. As a consequence, the model was able to reproduce the absorption lines associated with a slow outflow component with a velocity of approximately $100 \ \mathrm{km} \ \mathrm{s}^{-1}$. However, the model did not reproduce the blue-shifted absorption lines corresponding to a fast outflow component with a velocity of approximately $1000 \ \mathrm{km} \ \mathrm{s}^{-1}$. This would be natural because their hydrodynamic model is based on \cite{Wada2016}, in which there is no fast-velocity gas with $\simeq 1000 \ \mathrm{km} \ \mathrm{s}^{-1}$ even at the innermost radius of $0.125 \ \mathrm{pc}$. \cite{Ogawa2022} confirmed that if such a fast velocity component presents in the central sub-parsec region, the absorption features can be reproduced. Therefore, a high-resolution, physics-based model for the central sub-parsec-scale region is necessary for further studies.

In this study, we performed X-ray radiative transfer calculations based on three-dimensional radiative hydrodynamic simulations for the sub-parsec-scale region surrounding the SMBH with a mass of $M_{\mathrm{BH}} = 2.8 \times 10^{7} M_{\odot}$. The photoionization equilibrium of the circumnuclear gas irradiated by the AGN was calculated. We then performed the three-dimensional X-ray radiative transfer calculations, in which absorption and scattering of photons are fully taken into account. We applied our model to the X-ray spectrum of NGC~3783 observed with the Chandra HETG. The remainder of this paper is organized as follows. In \hyperref[Section0200]{Section~2}, we present the radiative hydrodynamic simulation, the photoionized equilibrium calculations, and X-ray radiative transfer. In \hyperref[Section0300]{Section~3}, we describe the results of our simulations and apply our model to the X-ray spectrum of NGC~3783. In \hyperref[Section0400]{Section~4}, we discuss the origin of H-like iron absorption lines and neutral iron fluorescence emission lines. Finally, we present our conclusions in \hyperref[Section0500]{Section~5}.
\begin{figure*}\label{Figure0001}
\gridline{\fig{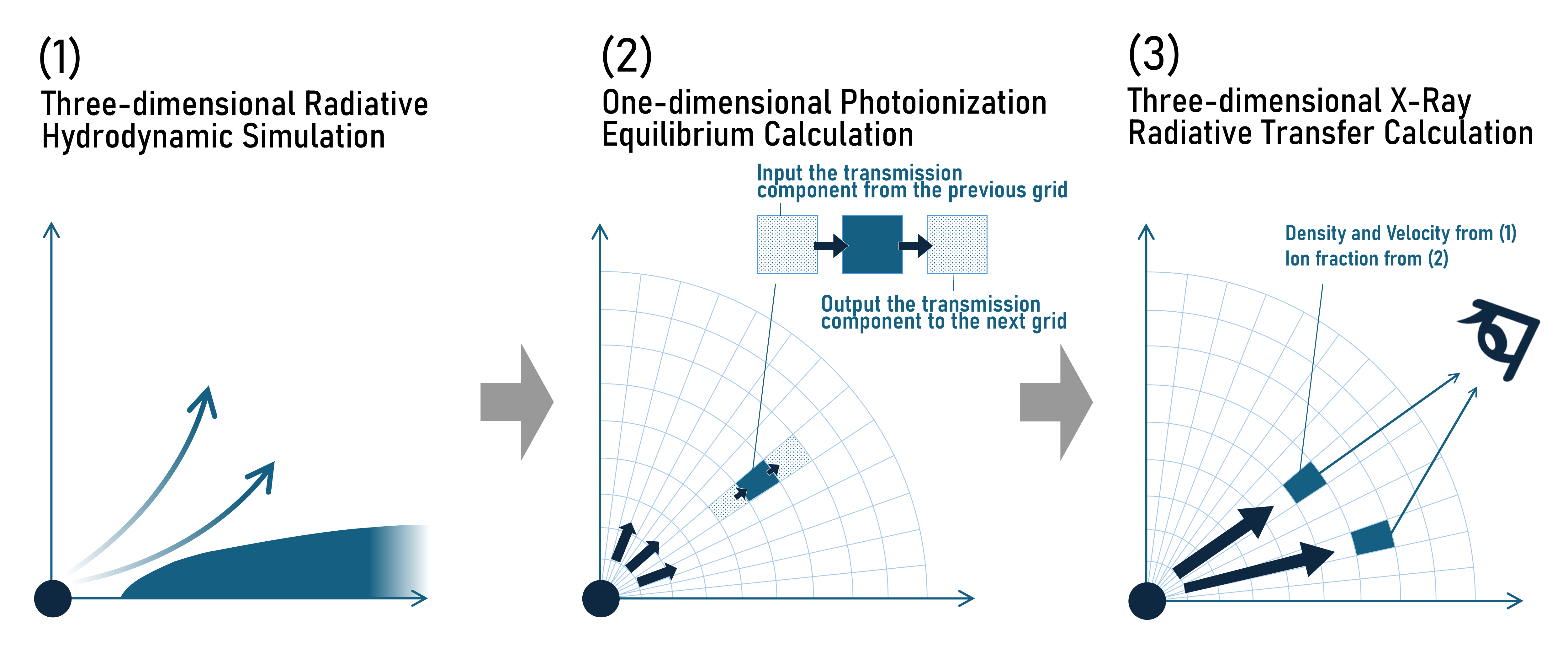}{0.8\textwidth}{}}
\caption{Our calculation procedures. (1) Three-dimensional radiative hydrodynamic simulation. (2) One-dimensional photoionization equilibrium calculations based on the radiative hydrodynamic simulation with the Cloudy code \citep{Ferland2017}. Here we performed photoionization equilibrium calculations from the inside to the outside for each of the inclination angles. (3) Monte Carlo X-ray radiative transfer calculations based on both radiative hydrodynamic simulations and photoionization equilibrium calculations with the MONACO code \citep{Odaka2016}.}
\end{figure*}\clearpage
\section{Methods}\label{Section0200}
\hyperref[Figure0001]{Figure~1} provides an illustration of the calculation procedure. The calculations consist of three steps. (1) Three-dimensional radiative hydrodynamic simulations \citep[][in prep.]{Wada2025}, (2) photoionization equilibrium calculations with the Cloudy photoionization code \citep{Ferland2017}, and (3) X-ray radiative transfer calculations with Monte Carlo simulations for astrophysics and cosmology (MONACO; \citealt{Odaka2011, Odaka2016}) code. \hyperref[Section0201]{Section~2.1} presents the basic equations and physical processes considered in the radiative hydrodynamic simulations. \hyperref[Section0202]{Section~2.2} describes the photoionization equilibrium calculations based on three-dimensional radiative hydrodynamic simulations. \hyperref[Section0203]{Section~2.3} presents the Monte Carlo X-ray radiative transfer calculations based on both radiative hydrodynamic simulations and photoionization equilibrium calculations.
\subsection{Radiative Hydrodynamic Simulations}\label{Section0201}
The numerical method is based on \cite{Wada2012}, which investigates the dynamics of the dusty gas irradiated by the AGN on a parsec scale. The model has successfully reproduced radio \citep{Wada2018a, Izumi2018, Uzuo2021, Izumi2023, Baba2024}, infrared \citep{Wada2016, Matsumoto2023}, and optical observations \citep{Wada2018b, Wada2023} of nearby AGNs. In this study, we apply the method to the sub-parsec-scale region, taking into account the heating of cold, warm, and hot ionized gas by X-ray radiation. To simulate the three-dimensional evolution of a rotating gas disk within a fixed spherical gravitational potential influenced by radiation from the central source, we numerically solve the following equations.

\begin{align}
\frac{\partial \rho}{\partial t} + \nabla \cdot (\rho \bm{v})                               & = 0,                          \label{Equation01}\\
\frac{\partial (\rho \bm{v})}{{\partial t}} + \rho (\bm{v}\cdot \nabla)\bm{v}+{\nabla p}    & = 
-\rho \nabla \Phi_{\mathrm{SMBH}} + \rho \bm{f}_{\mathrm{rad}}^{r},                                                         \label{Equation02}\\
\frac{\partial (\rho E)}{\partial t} + \nabla \cdot [(\rho E+p)\bm{v}]                      & = 
\rho \bm{v} \cdot \nabla \Phi_{\mathrm{SMBH}} + \rho \bm{v} \cdot \bm{f}_{\mathrm{rad}}^{r}
+ \rho \Gamma_{\mathrm{UV}}(G_0) + \rho \Gamma_{\mathrm{X}} - \rho^{2} \Lambda(T_{\mathrm{gas}})                            \label{Equation03}
\end{align}

In these equations, $\rho$ represents the gas density, $\bm{v}$ is the gas velocity, $p$ is the gas pressure, $\Phi_{\mathrm{SMBH}}$ is the potential of the SMBH, $\bm{f}_{\mathrm{rad}}^{r}$ is the radiative force, $E$ is the specific total energy, $\Gamma_{\mathrm{UV}}$ is the heating due to the photoelectric effect, $\Gamma_{\mathrm{X}}$ is the X-ray heating rate, and $\Lambda$ is the cooling function. 

In the momentum conservation equation \hyperref[Equation02]{Equation~2}, we adopted the following time-independent potentials of the SMBH $\Phi_{\mathrm{SMBH}}(r)$
\begin{equation}
\Phi_{\mathrm{SMBH}}(r) = -\frac{GM_{\mathrm{SMBH}}}{(r^2 + b^2)^{\frac{1}{2}}}
\end{equation}
where $M_{\mathrm{BH}} = 2.80 \times 10^7 M_{\odot}$. To avoid numerical errors and small time steps near the innermost numerical grid cells, the gravitational potential is smoothed for $b = 2.5 \times 10^{-3} \ \mathrm{pc}$. Although we assumed an Eddington ratio of approximately $10\%$ in this calculation, the effective Eddington ratio is greater than this value. This is because the gravitational potential of the SMBH is effectively reduced as a result of the gravitational softening. We also examined the radial component of the radiation pressure:
\begin{align}
\bm{f}_{\mathrm{rad}}(r) & = \int \frac{\chi_{\mathrm{T}}\bm{F}(r)_{\nu}}{c} d\nu,                         \\
\bm{F}_{\nu}(r) & = \frac{L_{\nu}(\theta) e^{-\tau_{\mathrm{T}}}}{4\pi r^{2}}\bm{e}_{r}.
\end{align}
Here, $\chi_{\mathrm{T}}$ represents the total mass extinction coefficient due to Thomson scattering and $\bm{F}_{\nu}(r)$ is the radial component of the flux at a given radius with $\tau_{\rm T} =\int \chi_{\rm T} \rho ds$. The optical depth $\tau_{\rm T}$ is computed at each time step along a ray originating from the central source at every grid point; that is, $256^{3}$ rays are traced within the computational domain.

In the energy conservation equation \hyperref[Equation03]{Equation~3}, we used a cooling function derived from a radiative transfer model of photodissociation regions \citep{Meijerink2005}. The heating contribution of the photoelectric effect $\Gamma_{\mathrm{UV}}$ is taken into account in the model. It is assumed that a uniform far-ultraviolet (FUV) radiation field with a strength of $G_0 = 10^4$ originates from star-forming regions within the central several tens of parsecs. We note that the photoelectric heating does not significantly affect the thermal structure of the gas in the present model.

Finally, we account for X-ray heating, including interactions between high-energy secondary electrons and thermal electrons (Coulomb heating), as well as $H_{2}$ ionization heating in both the cold and the warm gas phases \citep{Maloney1996, Meijerink2005}. The Coulomb heating rate $\Gamma_{\mathrm{X,c}}$ for gas with a number density $n$ is given by
\begin{equation}
\Gamma_{\mathrm{X,c}} = \eta n H_{\mathrm{X}},
\end{equation}
where $\eta$ represents the heating efficiency \citep{Meijerink2005}, $H_{\mathrm{X}} = 3.8 \times 10^{-25} \xi \ \mathrm{ergs} \ \mathrm{s}^{-1}$ is the X-ray energy deposition rate, and the ionization parameter $\xi = 4\pi F_{\mathrm{X}}/n  = L_{\mathrm{X}} e^{-\int \tau_{\nu} d\nu}/n r^{2}$ with $L_{\mathrm{X}} = 0.1 L_{\mathrm{AGN}}$. For an optically thin hot gas with $T \gtrsim 10^{4} \ \mathrm{K}$,the effects of Compton heating and X-ray photoionization heating are included. The net heating rate \citep{Blondin1994} is approximately 
\begin{equation}
\Gamma_{\mathrm{X,h}} = 8.9\times 10^{-36} \xi (T_{\mathrm{X}} - 4T) +1.5\times 10^{-21} \xi^{1/4} T^{-1/2}(1-T/T_{\mathrm{X}}) \ \mathrm{erg} \ \mathrm{s}^{-1} \ \mathrm{cm}^{3},
\end{equation}
with the characteristic temperature of the bremsstrahlung radiation $T_{\mathrm{X}} = 10^{8} \ \mathrm{K}$.

We solve the basic equations using the advection upstream splitting method \citep{Liou1993, Wada2001}. The simulations utilize a grid of $256^3$ points. A uniform Cartesian grid spans a region of $0.16 \ \mathrm{pc} \times \ 0.16 \ \mathrm{pc} \times \ 0.08 \ \mathrm{pc}$ around the galactic center. The spatial resolution for the $x$ and $y$ directions is $6.250 \times 10^{-4} \ \mathrm{pc}$ and that for the $z$ direction is $3.125 \times 10^{-4} \ \mathrm{pc}$. We assumed the $2$--$10 \ \mathrm{keV}$ X-ray luminosity of $5.0 \times 10^{42} \ \mathrm{erg} \ \mathrm{s}^{-1}$. To establish a quasi-steady initial condition without radiation feedback, we first simulate an axisymmetric, rotationally supported thin disk with a uniform density profile, a thickness of $6.250 \times 10^{-4} \ \mathrm{pc}$, and a total gas mass of $M_{\mathrm{gas}} = 6.68 \times 10^{6} M_{\odot}$. As boundary conditions for inner regions, we set physical quantities such as the mass density $\rho$ and the velocity constant ($\rho = 10 M_{\sun} \ \mathrm{pc}^{-3}$ and $v = 0 \ \mathrm{km} \ \mathrm{s}^{-1}$) for regions with radii smaller than $2.5 \times 10^{-4} \ \mathrm{pc}$. As boundary conditions for the outer regions, we set outflow boundary conditions.
\subsection{Photoionization Equilibrium Calculations}\label{Section0202}
We performed photoionization equilibrium calculations using the Cloudy code \citep{Ferland2017} based on three-dimensional radiative hydrodynamic simulations. The Cloudy code enables the calculation of the physical conditions of the photoionized gas. Since Cloudy is a one-dimensional photoionization equilibrium calculation code, it is necessary to transform the results of the radiative hydrodynamic simulations from a Cartesian coordinate system to a spherical coordinate system. The radial range from $0.00$ to $0.02 \ \mathrm{pc}$ was divided into $40$ grids, and the polar angular range from $0$ to $\pi/2$ was divided into $40$ grids.

In this study, we performed one-dimensional photoionization equilibrium calculations based on radiative hydrodynamic simulations from the inner to the outer regions, for four values of $\phi$. Although the observed $2$--$10 \ \mathrm{keV}$ luminosity of NGC~3783 is approximately $1.2 \times 10^{43} \ \mathrm{erg} \ \mathrm{s}^{-1}$ \citep{Mao2019}, we assumed $5.0 \times 10^{42} \ \mathrm{erg} \ \mathrm{s}^{-1}$ as the initial condition. This is because we prioritized the consistency with the assumption of radiative hydrodynamic simulations. If the observed $2$--$10 \ \mathrm{keV}$ luminosity is adopted, the overall ionization parameter may be approximately factor 2 larger. Furthermore, the photon index of $1.60$ was assumed, based on observations of NGC~3783 \citep{Mao2019}. We then performed the photoionization equilibrium calculations from the inner to outer regions, using the output X-ray spectra from the previous grid as the input spectra for the next. Although the Cloudy calculation further divides each grid into several shells, it is not possible to consider all of these areas due to computer memory limitations. The ionic fraction was simply averaged over the ionic fraction of each divided shell. The solar abundances of \cite{Anders1989} were used.
\subsection{X-Ray Radiative Transfer}\label{Section0203}
To compute the X-ray spectra based on the radiative hydrodynamic simulation, we used the Monte Carlo Simulation for Astrophysics and Cosmology (MONACO; \citealt{Odaka2011, Odaka2016}) code version 1.8.0. This code employs the Geant4 framework \citep{Agostinelli2003, Allison2006, Allison2016} to track photons in complex geometric structures. Although Geant4 implements physical processes, MONACO uses optimized physical processes for astrophysics. MONACO incorporates three sets of physical processes: (1) X-ray reflection from neutral matter \citep{Odaka2011, Furui2016, Tanimoto2019, Tanimoto2022, Tanimoto2023, Uematsu2021}, (2) Comptonization in a hot flow \citep{Odaka2013, Odaka2014}, and (3) photon interactions in a photoionized plasma \citep{Watanabe2006, Hagino2015, Hagino2016, Tomaru2018, Tomaru2020, Mizumoto2021}.

In this simulation, we modeled photon interactions in photoionized plasmas. We performed X-ray radiative transfer calculations using a total of 64,000 grids. These included $40$ grids for the radial range from $0.00 \ \mathrm{pc}$ to $0.02 \ \mathrm{pc}$, $40$ grids for the angular range from $0$ to $\pi/2$, and $40$ grids for the azimuthal angle range from $0$ to $2\pi$. We note that although radiative hydrodynamic simulations ranged from $0.00 \ \mathrm{pc}$ to $0.08 \ \mathrm{pc}$, we confirmed that almost all absorption lines are produced the dense gas inside $r = 0.02 \ \mathrm{pc}$, and the relatively diffuse gas between $0.02 \ \mathrm{pc}$ and $0.08 \ \mathrm{pc}$, which is the outer boundary of the computational box, does not affect the X-ray spectrum. Each grid contained the hydrogen number density and three-dimensional velocities obtained from the radiative hydrodynamic simulations, as well as the ion fractions derived from the photoionization equilibrium calculations. In this calculation, we considered $13$ elements, including H, He, C, N, O, Ne, Mg, Si, S, Ar, Ca, Fe, and Ni. We also assumed a solar abundance of \cite{Anders1989}. Since we are mainly interested in absorption and emission lines ranging from $6 \ \mathrm{keV}$ to $7 \ \mathrm{keV}$, we included iron and nickel from H-like to Ne-like ions, and other elements from H-like to He-like ions. A total of four billion photons were generated from a source at the origin, with energies ranging from $2 \ \mathrm{keV}$ to $10 \ \mathrm{keV}$ and a photon index of $1.60$.\clearpage
\begin{figure}\label{Figure0002}
\gridline{\fig{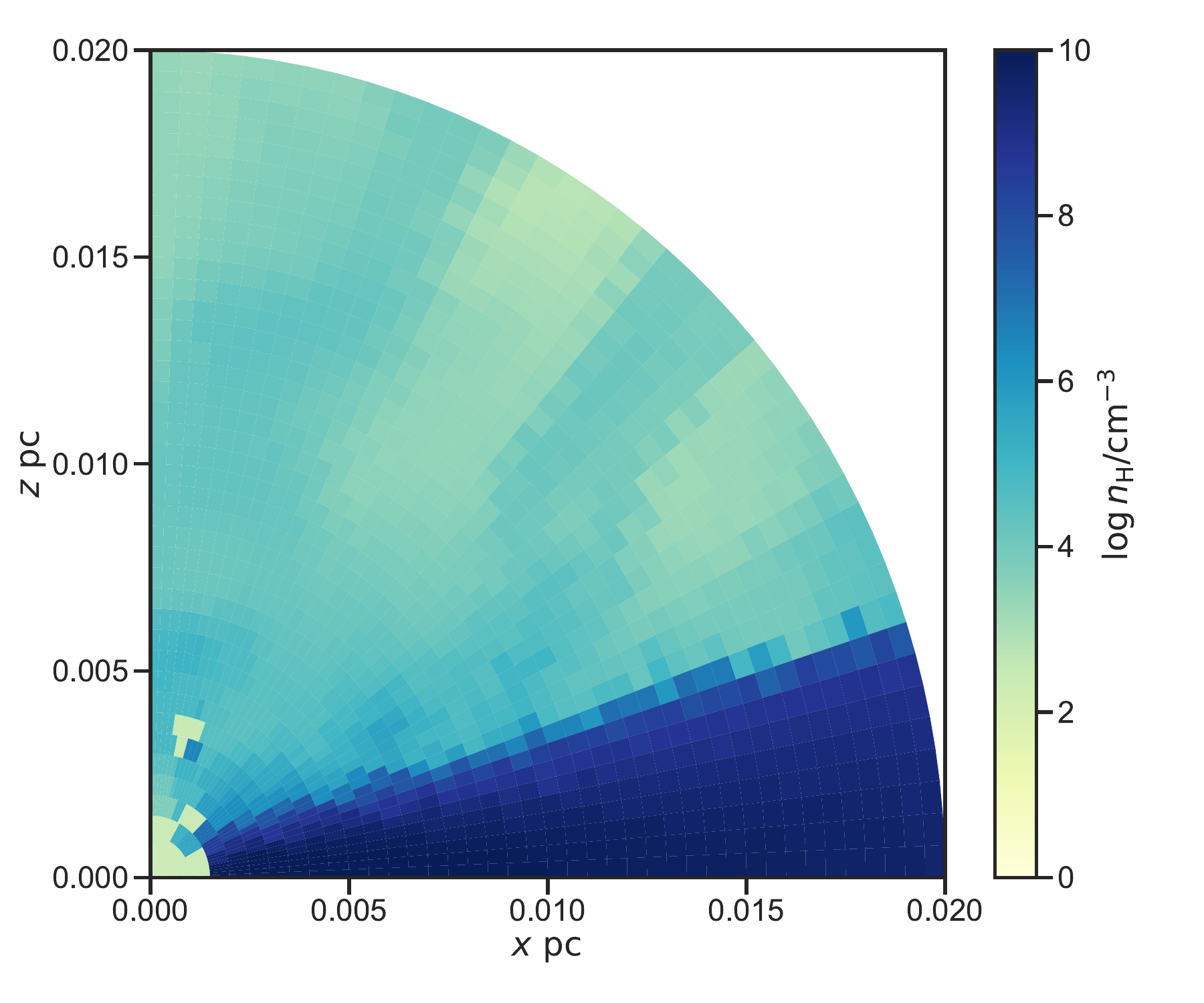}{0.5\textwidth}{(a)}}
\gridline{\fig{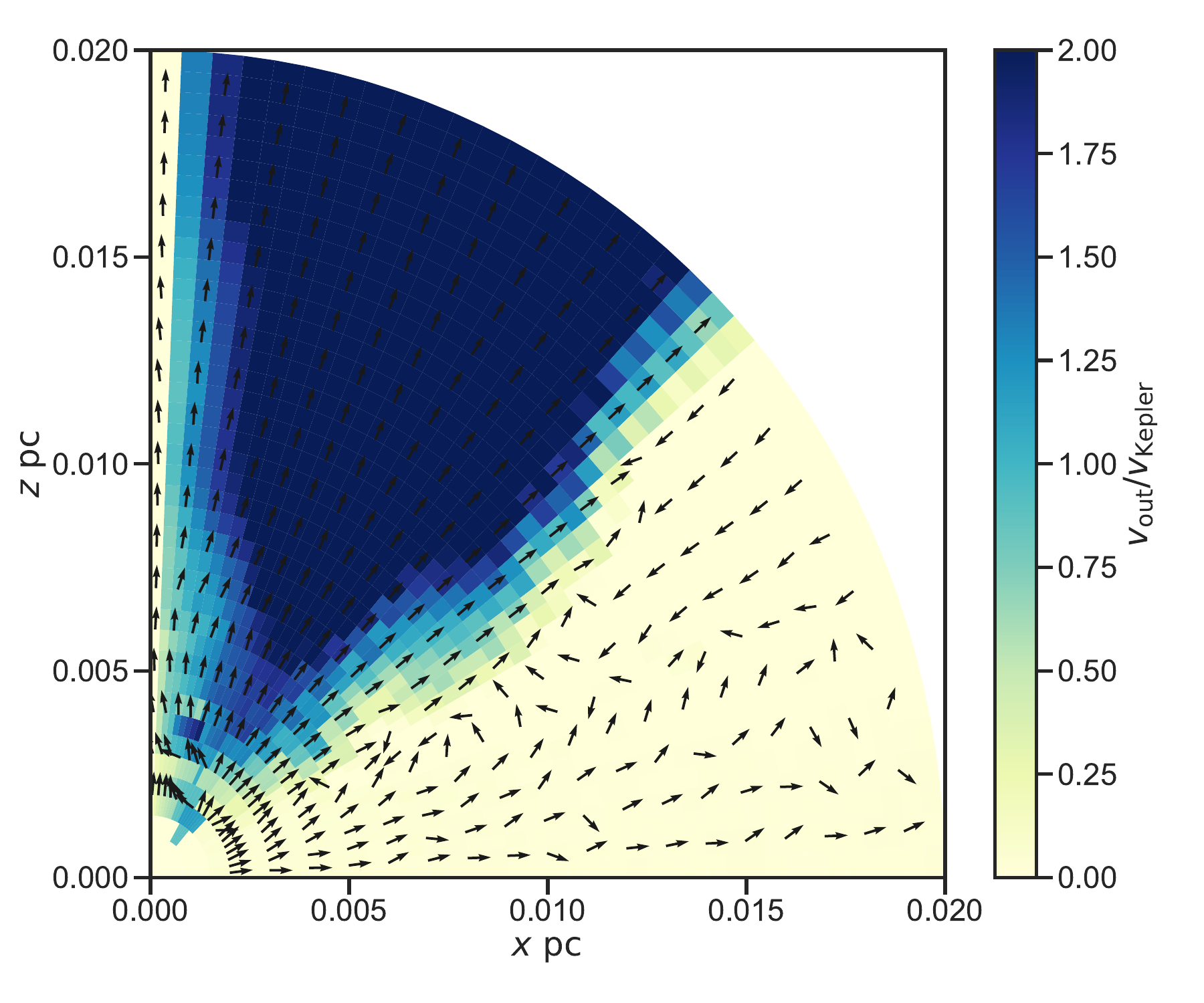}{0.5\textwidth}{(b)}\fig{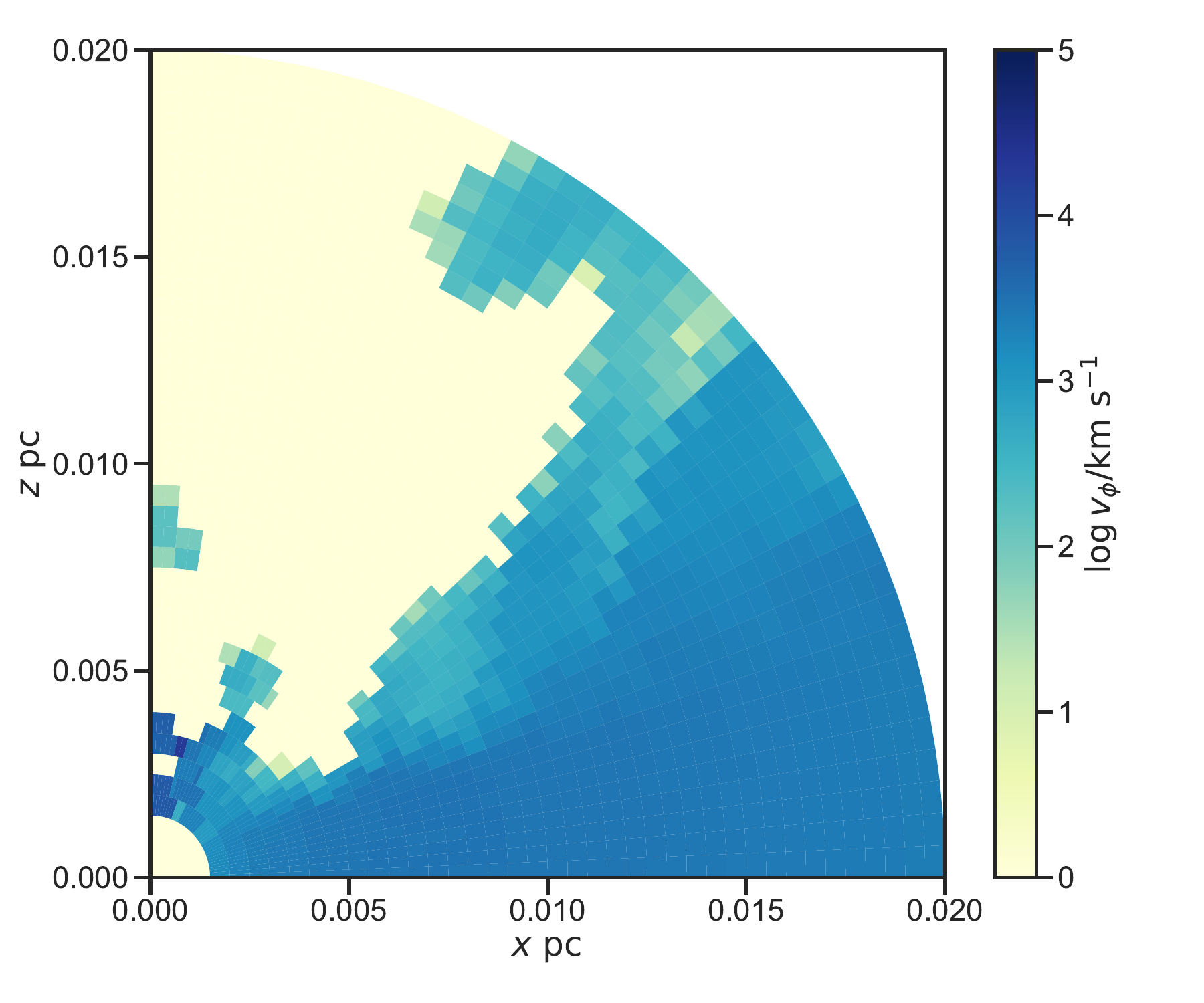}{0.5\textwidth}{(c)}}
\caption{(a) The distribution of the hydrogen number density $\log n_{\mathrm{H}}/\mathrm{cm}^{-3}$. (b) The distribution of the ratio of the radial velocity to Keplerian velocity $v_{\mathrm{out}}/v_{\mathrm{Kepler}}$. The black arrows represent the unit vector of velocity in the grid. (c) The distribution of the azimuthal velocity $\log v_{\phi}/\mathrm{km} \ \mathrm{s}^{-1}$.}
\end{figure}\clearpage
\section{Results}\label{Section0300}
We present the results obtained from our calculations. \hyperref[Section0301]{Section~3.1} shows the distribution of the hydrogen number density, the radial outflow velocity, and the azimuthal velocity obtained from the radiative hydrodynamic simulations. In \hyperref[Section0302]{Section~3.2}, we present the distribution of the temperature and the ionization parameter inferred from the photoionized equilibrium calculations. \hyperref[Section0303]{Section~3.3} shows the transmitted component and the scattered component obtained from the X-ray radiative transfer calculations. In \hyperref[Section0304]{Section~3.4}, the X-ray spectral model obtained from the calculation is applied to the X-ray spectrum of NGC~3783 observed with the Chandra HETG.
\subsection{Results of Radiative Hydrodynamic Simulations}\label{Section0301}
\subsubsection{Distribution of Hydrogen Number Densities}
\hyperref[Figure0002]{Figure~2(a)} presents the distribution of the hydrogen number density $\log n_{\mathrm{H}}/\mathrm{cm}^{-3}$ for the azimuthal angle $\phi = 0\degr$. \hyperref[Figure0002]{Figure~2(a)} indicates that the distribution of the hydrogen number density changes at an inclination angle of approximately $70\degr$. If the angle of inclination is less than $70\degr$, the hydrogen number density is $10^{4}$--$10^{6} \ \mathrm{cm}^{-3}$. When the angle of inclination is greater than $70\degr$, the hydrogen number density is $10^{8}$--$10^{10} \ \mathrm{cm}^{-3}$.

\subsubsection{Distribution of Radial Outflow Velocity}
To investigate the dynamics in detail, we then examined the distribution of the radial velocities. \hyperref[Figure0002]{Figure~2(b)} presents the distribution of the ratio of radial velocity to Keplerian velocity $v_{\mathrm{out}}/v_{\mathrm{Kepler}}$. Here, black arrows represent the unit vector of velocity in the grid. \hyperref[Figure0002]{Figure~2(b)} indicates that the dynamics can be classified into three types. (1) If the angle of inclination is less than $50\degr$, the outflow is driven at speeds greater than approximately $1000 \ \mathrm{km} \ \mathrm{s}^{-1}$. (2) When the angle of inclination ranges from $50\degr$ to $70\degr$, the inner region is the outflow with velocities of $100$--$1000 \ \mathrm{km} \ \mathrm{s}^{-1}$, while the outer region is the inflow. (3) If the angle of inclination is greater than $70\degr$, the radial velocity is less than approximately $100 \ \mathrm{km} \ \mathrm{s}^{-1}$, and the Keplerian rotation is dominant.

We can estimate the mass outflow rate based on the hydrogen number density and the outflow velocity. The mass outflow rate $\dot{M}_{\mathrm{out}}$ is defined by the following equation
\begin{equation}
\dot{M}_{\mathrm{out}} = 4 \pi m_{\mathrm{p}} r^2 \sum_{i = 1}^{40} n_{\mathrm{H}}^{i} v_{\mathrm{out}}^{i} d\theta^{i}
\end{equation}
where $m_{\mathrm{p}}$ is the proton mass, $r$ is the radius, $n_{\mathrm{H}}$ is the hydrogen number density, $v_{\mathrm{out}}$ is the outflow velocity, and $d\theta$ is the solid angle of the grid. We calculated the mass outflow rate $\dot{M}_{\mathrm{out}}$ at $r = 0.02 \ \mathrm{pc}$ based on the hydrogen number density and the outflow velocity obtained from radiative hydrodynamic simulations. As a result, the mass outflow rate $\dot{M}_{\mathrm{out}}$ is approximately $4.0 M_{\sun} \ \mathrm{yr}^{-1}$. This value is consistent with the mass outflow rate of NGC~3783 estimated from the XMM-Newton observation \citep{Behar2003}.

\subsubsection{Distribution of Azimuthal Velocity}
\hyperref[Figure0002]{Figure~2(c)} presents the distribution of azimuthal velocity $\log v_{\phi}/\mathrm{km} \ \mathrm{s}^{-1}$. \hyperref[Figure0002]{Figure~2(c)} indicates that the gas in the equatorial direction is rotating at approximately $5000 \ \mathrm{km} \ \mathrm{s}^{-1}$ due to the gravitational potential of $\sim3 \times 10^{7} M_{\sun}$ SMBH in the center. This rotation may influence the profile of the observed emission lines (\hyperref[Section0303]{Section~3.3}). \hyperref[Figure0002]{Figure~2(c)} also suggests that the azimuthal velocity is anisotropic. This is because the outflow is driven in the polar direction. We note that since these regions have low density (\hyperref[Figure0002]{Figure~2(a)}) and are almost completely ionized, they have almost no effect on the observed X-ray spectrum. \clearpage
\begin{figure}\label{Figure0003}
\gridline{\fig{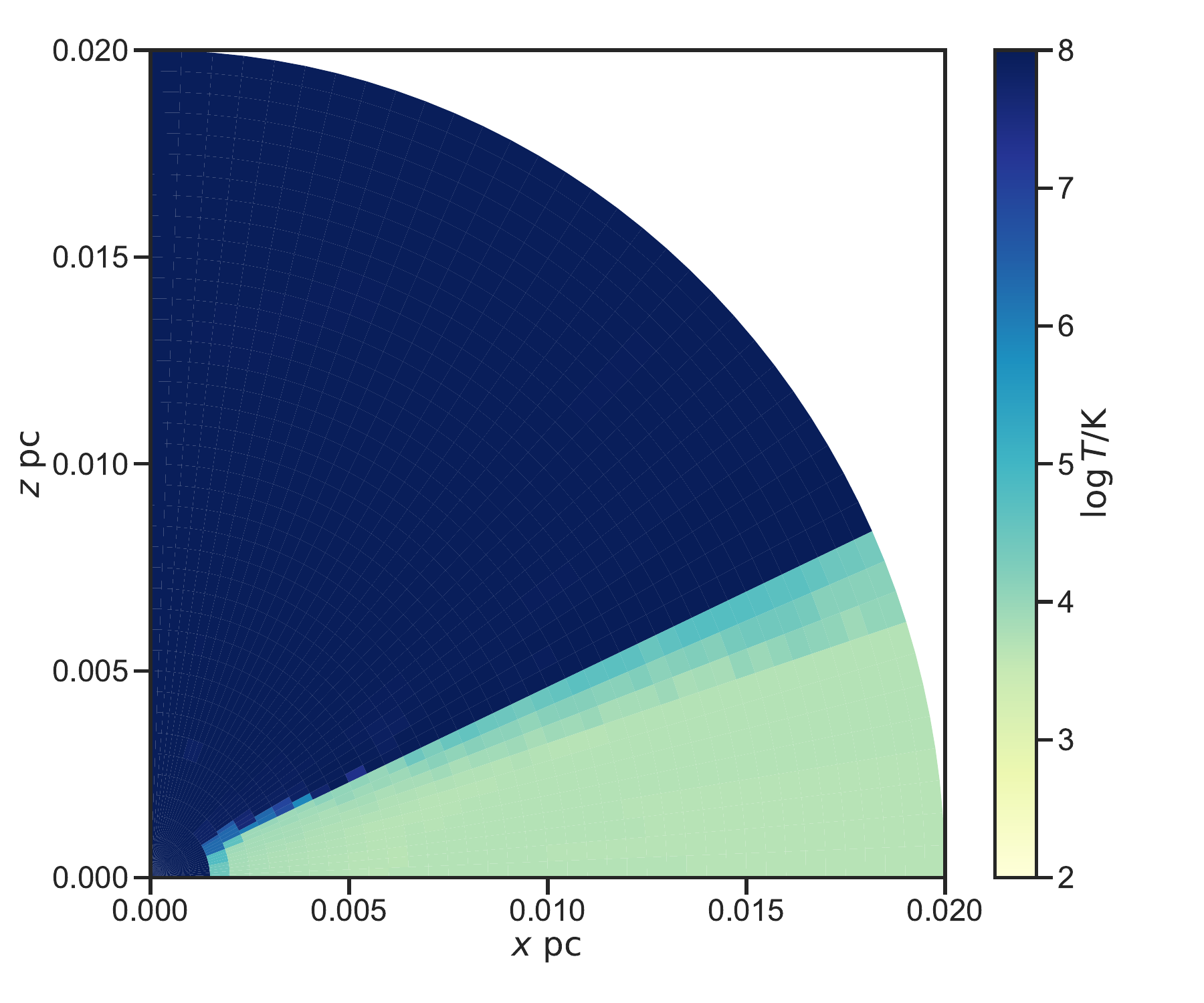}{0.5\textwidth}{(a)}{\fig{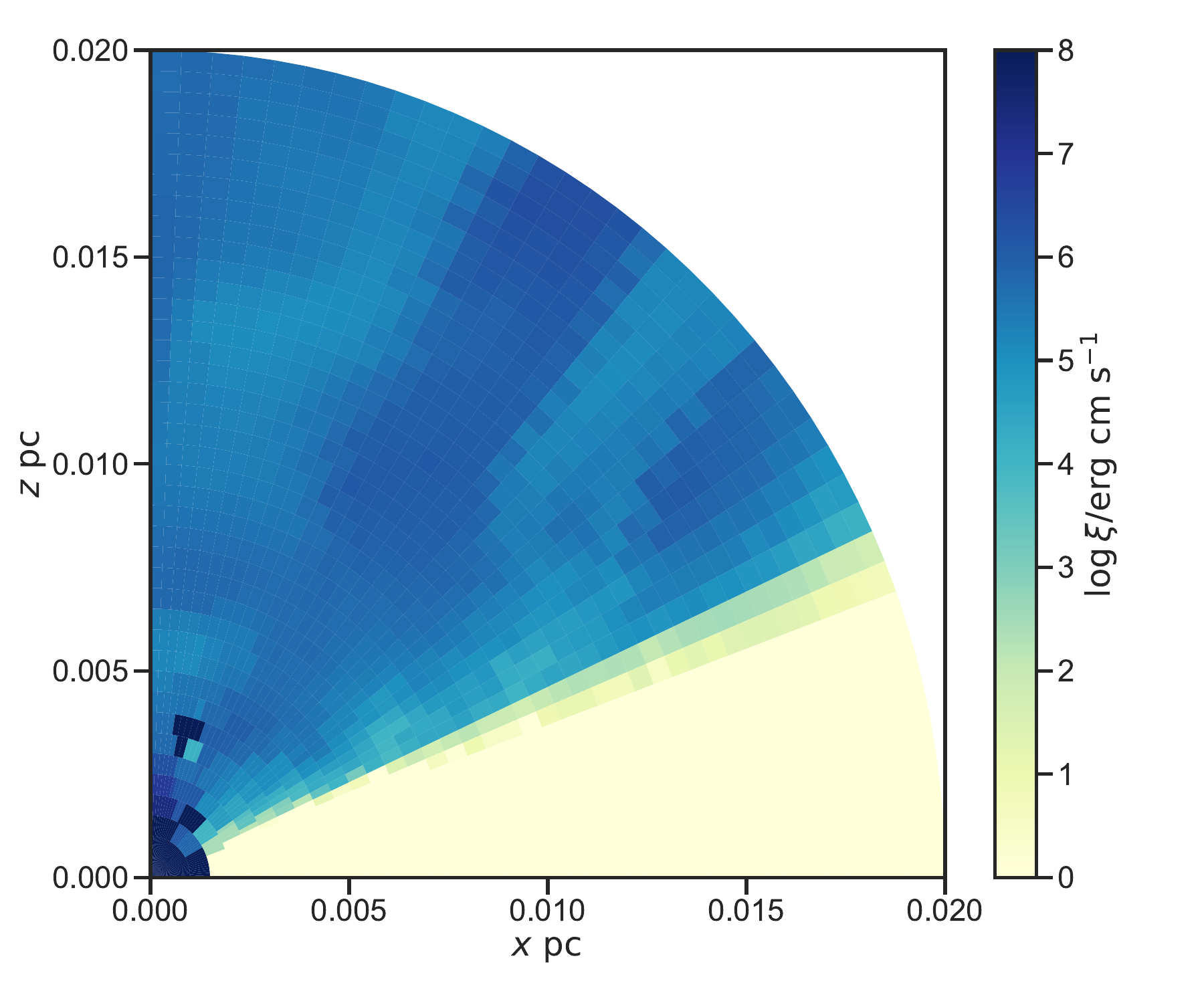}{0.5\textwidth}{(b)}}}
\caption{(a) The distribution of the temperature $\log T/\mathrm{K}$. (b) The distribution of the ionization parameter $\log \xi/\mathrm{erg} \ \mathrm{cm} \ \mathrm{s}^{-1}$.}
\end{figure}
\subsection{Results of Photoionized Equilibrium Calculations}\label{Section0302}
\subsubsection{Distribution of Temperature}
We performed the photoionized equilibrium calculation based on three-dimensional radiative hydrodynamic simulations. \hyperref[Figure0003]{Figure~3(a)} shows the distributions of the temperature $\log T/\mathrm{K}$. \hyperref[Figure0003]{Figure~3(a)} indicates that the temperature can also be classified into three types. (1) If the angle of inclination is less than $50\degr$, since the hydrogen number density is not high (\hyperref[Figure0002]{Figure~2(a)}), the temperature in the region is approximately $10^8 \ \mathrm{K}$ by the X-ray heating. (2) When the angle of inclination ranges from $50\degr$ to $70\degr$, the temperature is $10^5$--$10^6 \ \mathrm{K}$. (3) If the angle of inclination is greater than $70\degr$, the temperature in the region is approximately $10^4 \ \mathrm{K}$.

We compared the temperature distributions obtained from previous studies \citep{Holczer2007, Holczer2010, Holczer2012, Goosmann2016}. \cite{Goosmann2016} investigated the thermal instability of the warm absorber in NGC~3783 and proposed that gases with temperatures between $10^{4.5}$--$10^{5.0} \ \mathrm{K}$ are almost nonexistent due to the heating and cooling by M-shell iron. Similar temperature structures have been observed in sources such as IRAS~13349+2438 \citep{Holczer2007}, MCG~--06--30--15 \citep{Holczer2010}, and NGC~3516 \citep{Holczer2012}. The temperature distributions obtained from our calculations agree with these studies. The temperature distribution obtained from our calculations also shows few gases with temperatures between $10^{4.5}$--$10^{5.0} \ \mathrm{K}$.

\subsubsection{Distribution of Ionization Parameter}
\hyperref[Figure0003]{Figure~3(b)} shows the distributions of the ionization parameter $\log \xi/\mathrm{erg} \ \mathrm{cm} \ \mathrm{s}^{-1}$. If the angle of inclination is less than $50\degr$, the logarithmic ionization parameter $\log \xi/\mathrm{erg} \ \mathrm{cm} \ \mathrm{s}^{-1}$ is greater than $5$. This indicates that these regions are almost fully ionized. When the angle of inclination ranges from $50\degr$ to $70 \degr$, the logarithmic ionization parameter $\log \xi/\mathrm{erg} \ \mathrm{cm} \ \mathrm{s}^{-1}$ is $2$--$4$. If the angle of inclination is greater than $70\degr$, these regions are almost neutral. In other words, if the angle of inclination is approximately $50\degr$ to $70\degr$, absorption lines of photoionized iron can be observed (\hyperref[Section0301]{Section~3.1.1}).
\begin{figure}\label{Figure0004}
\gridline{\fig{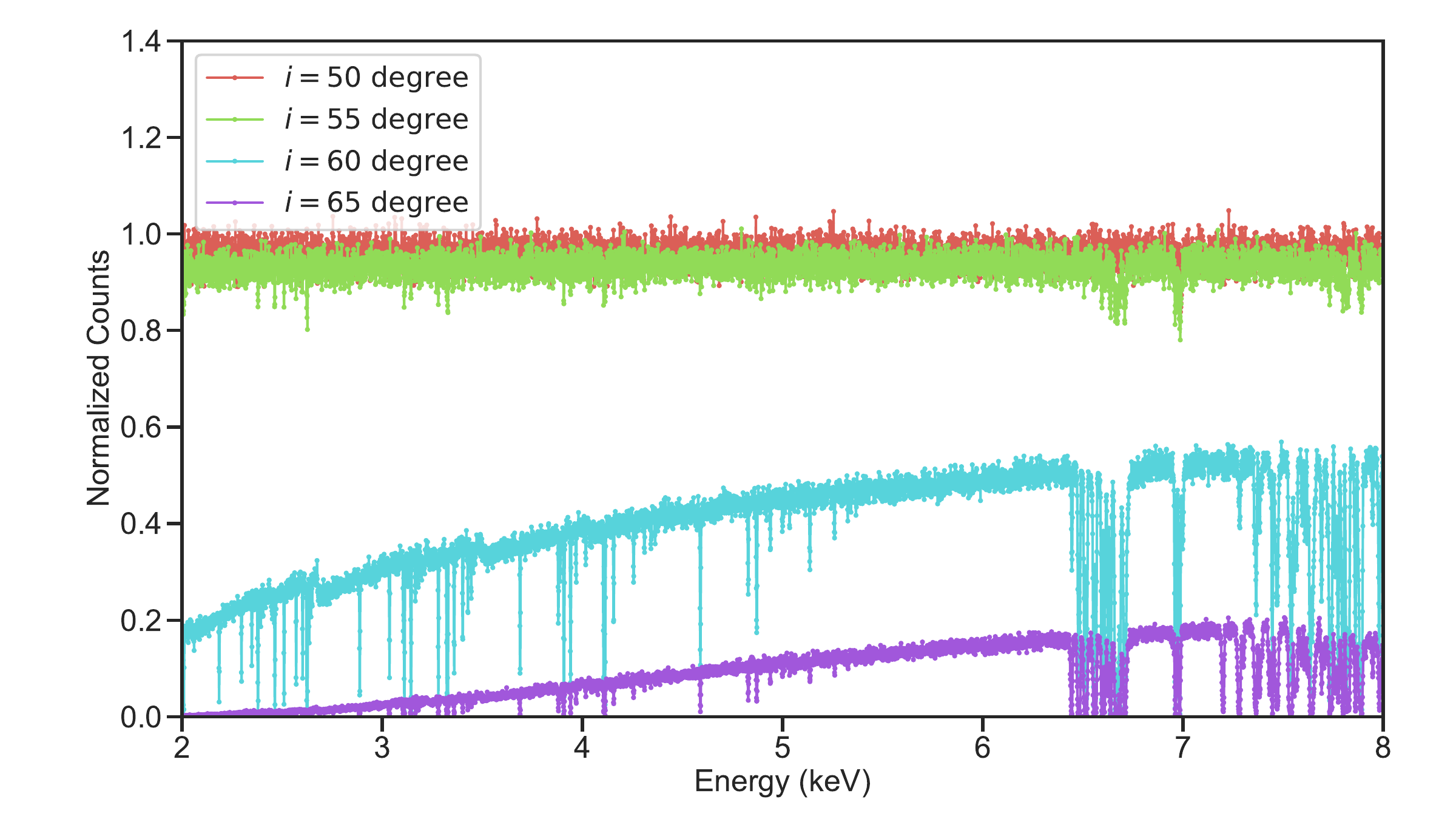}{0.5\textwidth}{(a)}\fig{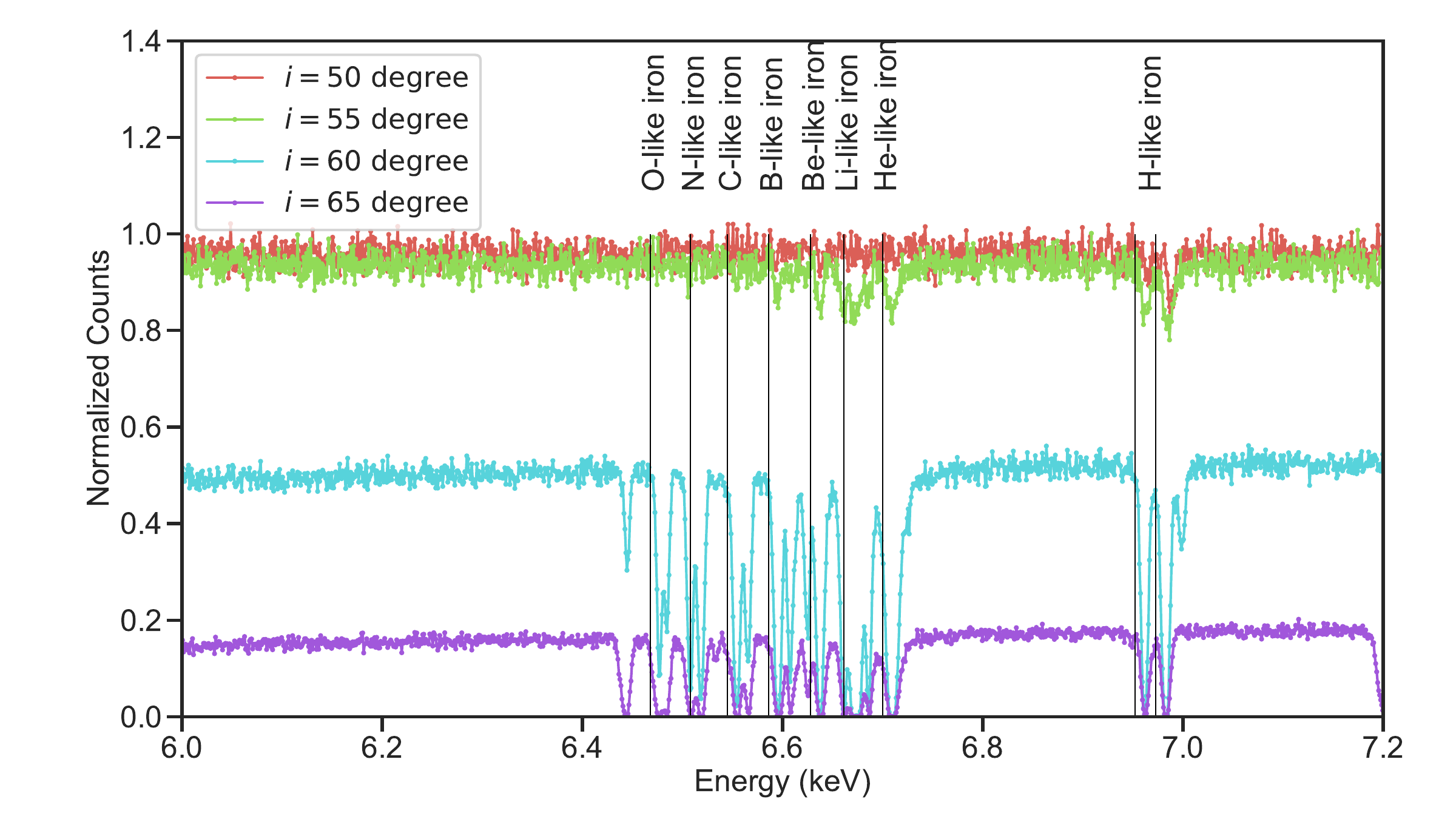}{0.5\textwidth}{(b)}}
\caption{(a) The transmitted component of the X-ray spectrum at $2$--$8 \ \mathrm{keV}$. (b) The transmitted component of the X-ray spectrum at $6.0$--$7.2 \ \mathrm{keV}$. The red, green, light blue, and purple lines show the X-ray spectrum at the angle of inclination $i = 50 \degr$, $55\degr$, $60\degr$, and $65\degr$, respectively. The black lines correspond to the energy of the absorption lines in the rest frame due to O-like iron ($6.468 \ \mathrm{keV}$), N-like iron ($6.508 \ \mathrm{keV}$), C-like iron ($6.545 \ \mathrm{keV}$), B-like iron ($6.586 \ \mathrm{keV}$), Be-like iron ($6.628 \ \mathrm{keV}$), Li-like iron ($6.661 \ \mathrm{keV}$), He-like iron ($6.700 \ \mathrm{keV}$), and H-like iron ($6.952 \ \mathrm{keV}$ and $6.973 \ \mathrm{keV}$). Energies of absorption lines are based on the Flexible Atomic Code \citep{Gu2008}.}
\end{figure}
\begin{figure}\label{Figure0005}
\gridline{\fig{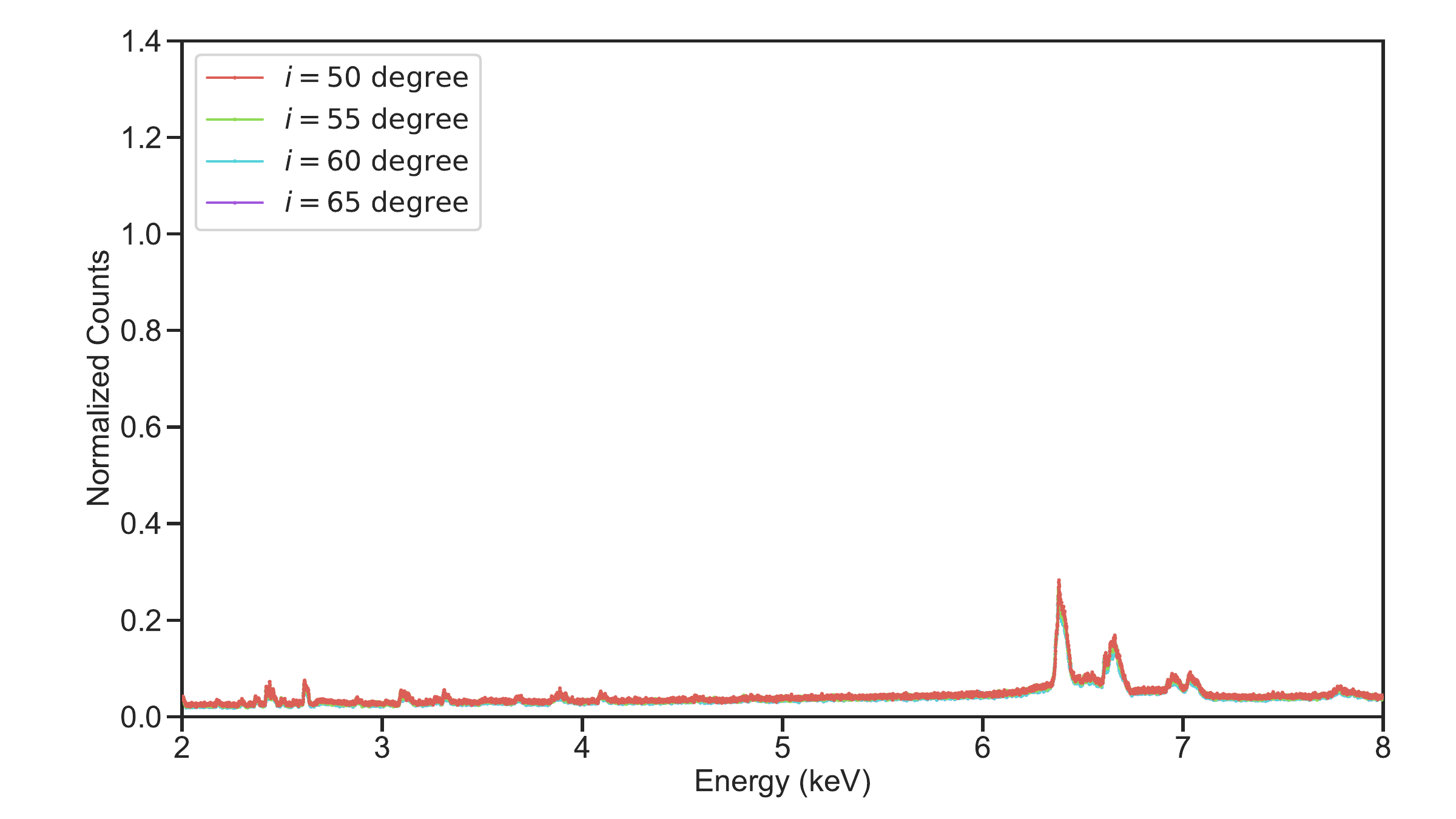}{0.5\textwidth}{(a)}\fig{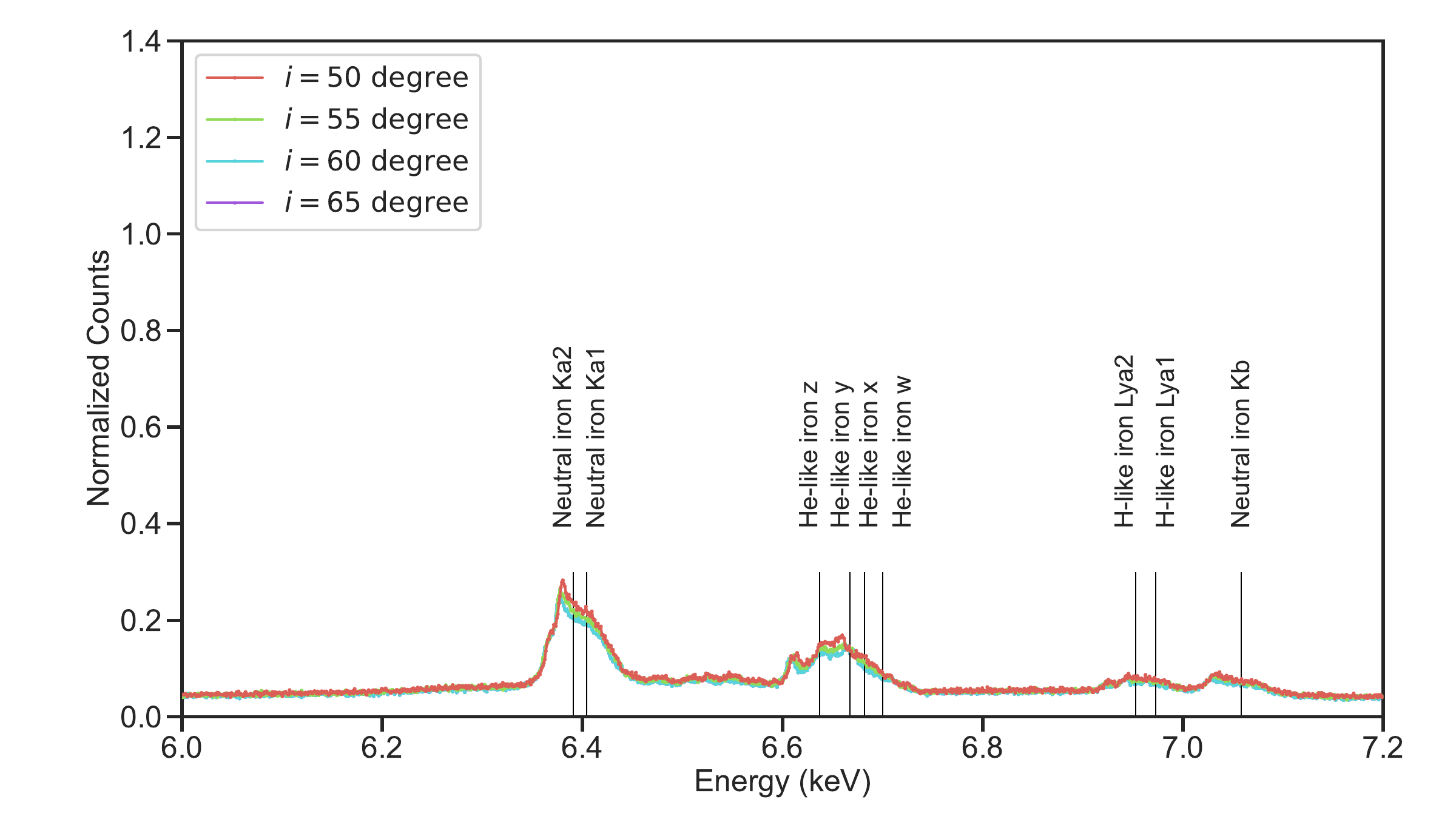}{0.5\textwidth}{(b)}}
\caption{(a) The scattered component of the X-ray spectrum at $2$--$8 \ \mathrm{keV}$. (b) The scattered component of the X-ray spectrum at $6.0$--$7.2 \ \mathrm{keV}$. The red, green, and light blue lines represent the X-ray spectrum at the inclination angle $i = 50\degr$, $55\degr$, $60\degr$, and $65\degr$, respectively. The black lines correspond to the energy of the emission lines in the rest frame due to neutral iron K$\alpha$2 ($6.391 \ \mathrm{keV}$), neutral iron K$\alpha$1 ($6.404 \ \mathrm{keV}$), He-like iron z ($6.637 \ \mathrm{keV}$), He-like iron y ($6.667 \ \mathrm{keV}$), He-like iron x ($6.682 \ \mathrm{keV}$), He-like iron w ($6.700 \ \mathrm{keV}$), H-like iron Ly$\alpha$2 ($6.952 \ \mathrm{keV}$), H-like iron Ly$\alpha$1 ($6.973 \ \mathrm{keV}$), and neutral iron K$\beta$ ($7.058 \ \mathrm{keV}$). Energies of absorption lines are based on the Flexible Atomic Code \citep{Gu2008}.}
\end{figure}
\begin{figure}\label{Figure0006}
\gridline{\fig{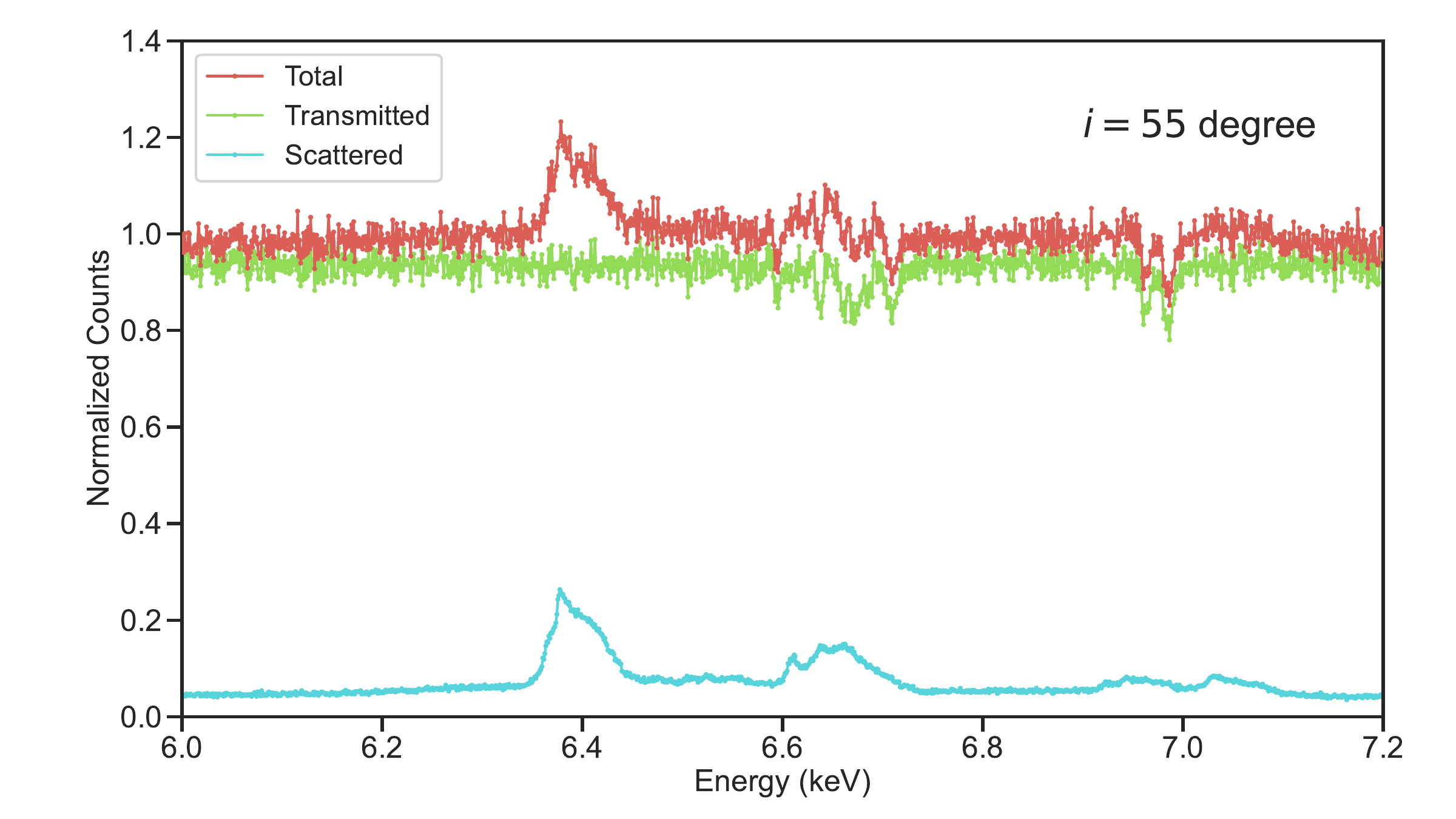}{0.5\textwidth}{(a)}\fig{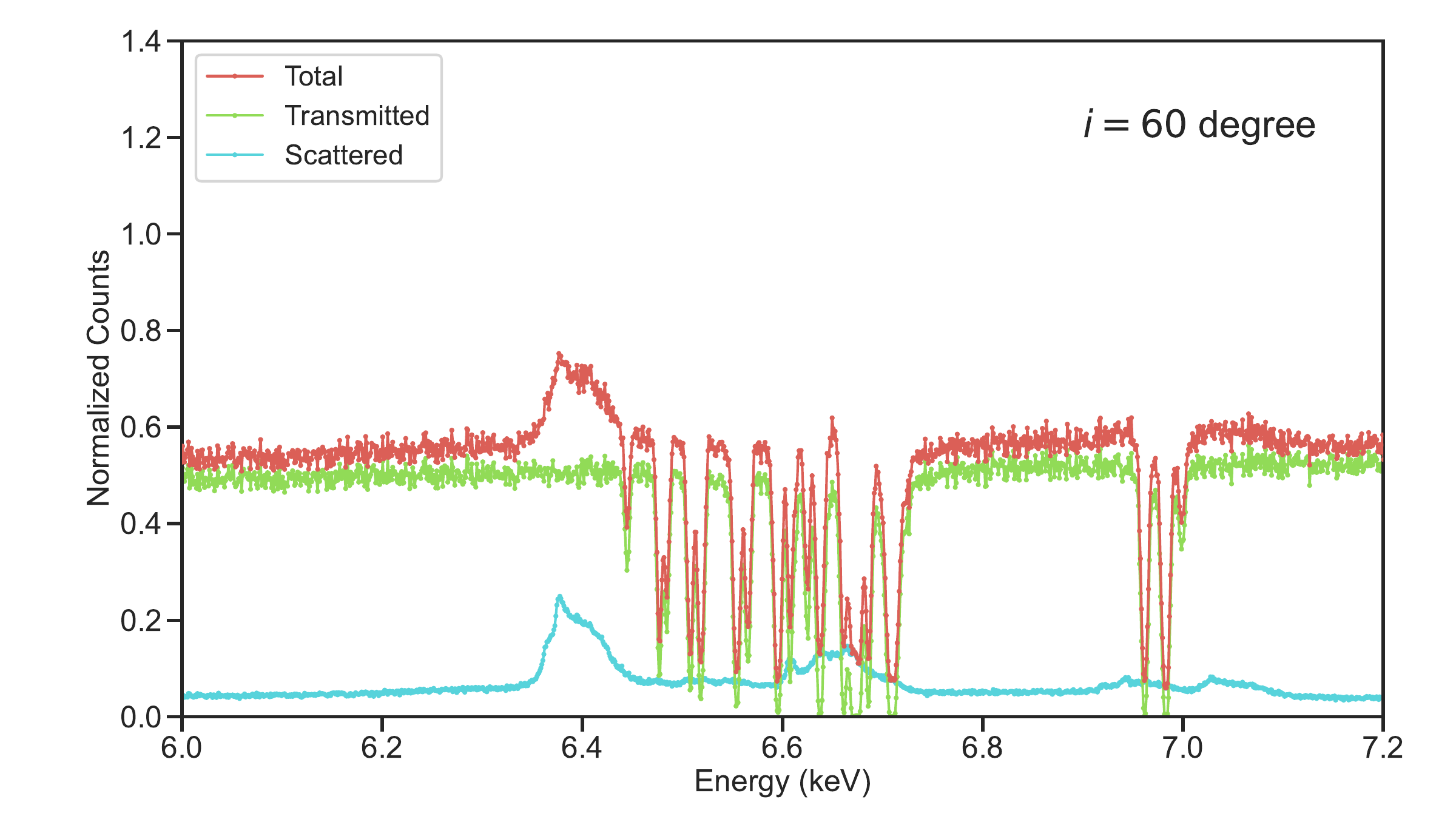}{0.5\textwidth}{(b)}}
\gridline{\fig{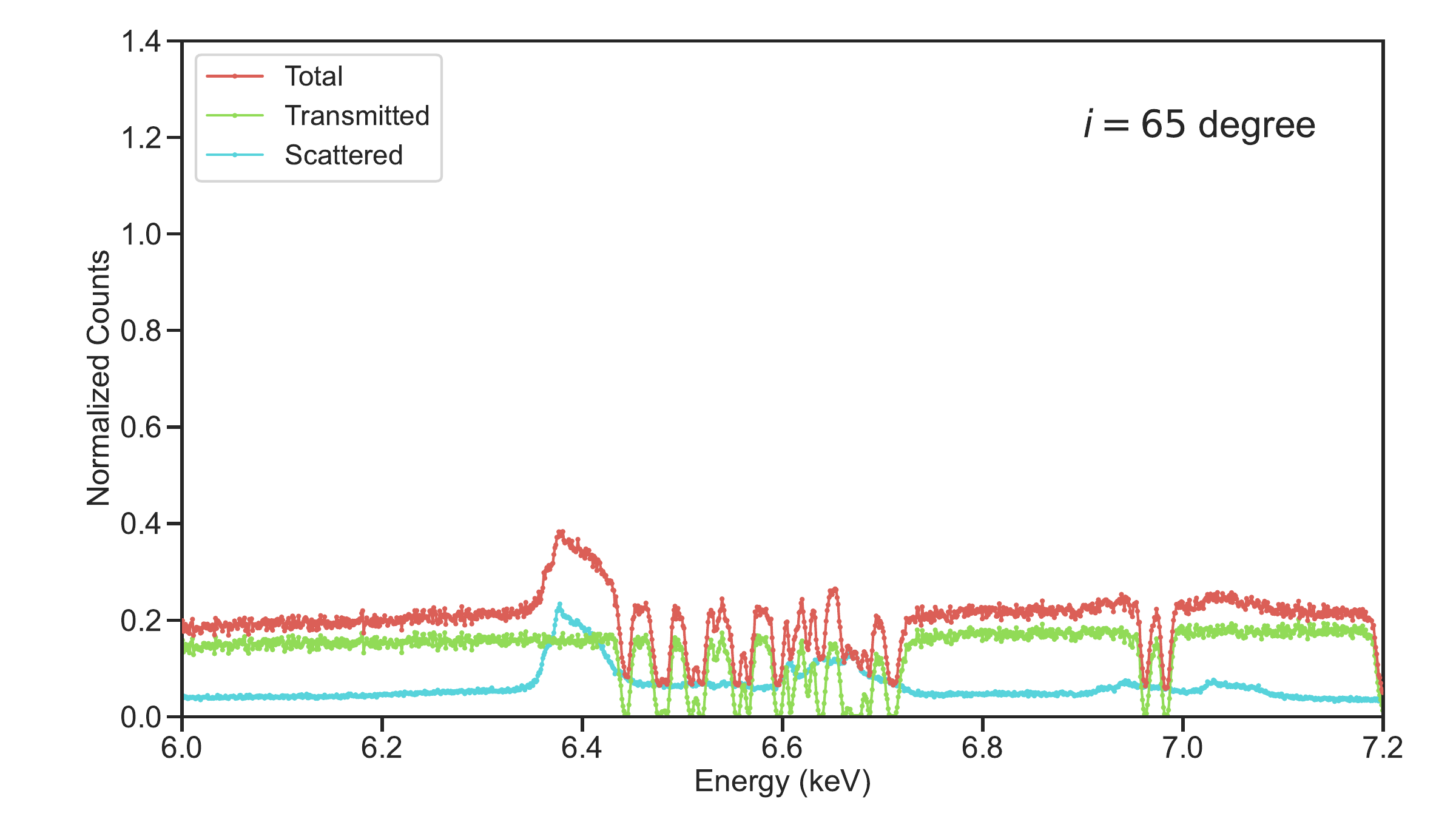}{0.5\textwidth}{(c)}\fig{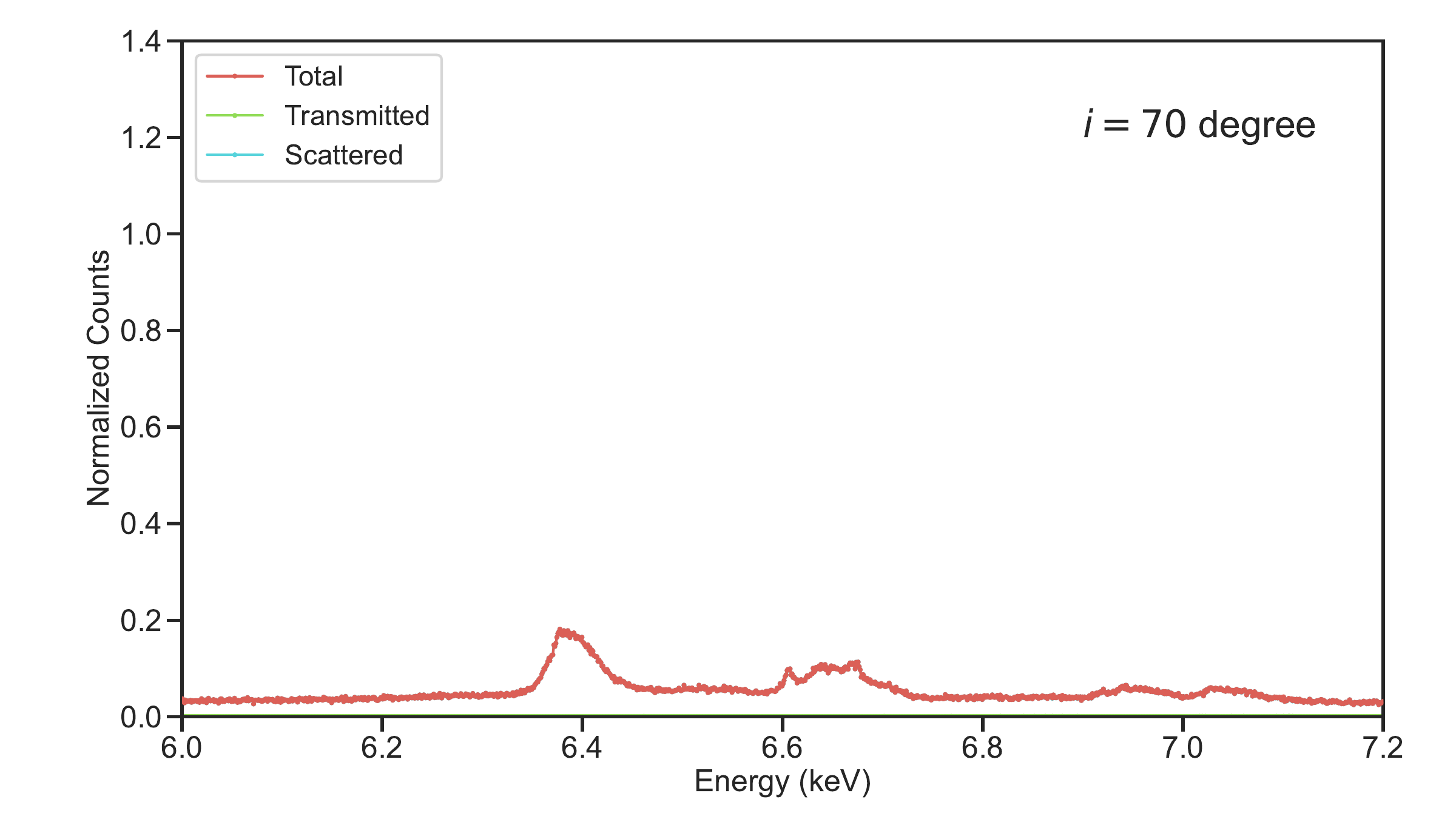}{0.5\textwidth}{(d)}}
\caption{(a) The total X-ray spectrum at the inclination angle at $i = 55\degr$. (b) The total X-ray spectrum at $i = 60\degr$. (c) The total X-ray spectrum at $i = 65\degr$. (d) The total X-ray spectrum at $i = 70\degr$. The red, green, light blue lines correspond to the total, the transmitted component, and the scattered component, respectively.}
\end{figure}\clearpage
\subsection{Results of X-Ray Radiative Transfer}\label{Section0303}
\subsubsection{Transmitted Component of the X-Ray Spectrum}
We computed the X-ray spectrum based on three-dimensional radiative hydrodynamic simulations. \hyperref[Figure0004]{Figure~4(a)} shows the dependence of the transmitted component of the X-ray spectrum at $2$--$8 \ \mathrm{keV}$ on the angle of inclination. Here, the energy resolution of the X-ray spectrum is $1 \ \mathrm{eV}$. If the angle of inclination is less than $50\degr$, the transmitted component shows almost no absorption lines. When the angle of inclination ranges from $55\degr$ to $65\degr$, the transmitted component exhibits absorption lines due to silicon, sulfur, argon, calcium, iron, and nickel. If the angle of inclination is greater than $70\degr$, the transmitted component is not observed due to the presence of Compton-thick material along the line of sight (\hyperref[Figure0002]{Figure~2a}).

Since our main interest is in absorption lines due to iron at $6$--$7 \ \mathrm{keV}$, \hyperref[Figure0004]{Figure~4(b)} shows the dependence of the transmitted component of the X-ray spectrum at $6.0$--$7.2 \ \mathrm{keV}$ on the angle of inclination. If the angle of inclination ranges from $55\degr$ to $65\degr$, the transmitted component shows the absorption lines of O-like iron ($6.468 \ \mathrm{keV}$), N-like iron ($6.508 \ \mathrm{keV}$), C-like iron ($6.545 \ \mathrm{keV}$), B-like iron ($6.586 \ \mathrm{keV}$), Li-like iron ($6.661 \ \mathrm{keV}$), He-like iron ($6.700 \ \mathrm{keV}$), H-like iron Ly$\alpha$2 ($6.952 \ \mathrm{keV}$), and H-like iron Ly$\alpha$1 ($6.973 \ \mathrm{keV}$). The observed energies of the absorption lines due to He-like iron and H-like iron are $6.711 \ \mathrm{keV}$, $6.963 \ \mathrm{keV}$, and $6.984 \ \mathrm{keV}$, respectively. In other words, these absorption lines are blue-shifted by a velocity of approximately $500 \ \mathrm{km} \ \mathrm{s}^{-1}$.

\subsubsection{Scattered Component of the X-Ray Spectrum}
Since MONACO is a three-dimensional X-ray radiative transfer code, it is capable of accurately accounting for the scattered component. \hyperref[Figure0005]{Figure~5(a)} shows the dependence of the scattered component of the X-ray spectrum at $2$--$8 \ \mathrm{keV}$. Here, we averaged the scattered components obtained from the four azimuthal angles $\phi = 0\degr$, $90\degr$, $180\degr$, and $270\degr$. \hyperref[Figure0005]{Figure~5(a)} indicates that the scattered component is almost independent of the angle of inclination compared to the transmitted component. This is because the scattered component includes photons scattered at various locations.

\hyperref[Figure0005]{Figure~5(b)} shows the dependence of the scattered component of the X-ray spectrum at $6.0$--$7.2 \ \mathrm{keV}$. The scattered component exhibits emission lines such as neutral iron K$\alpha$2 ($6.391 \ \mathrm{keV}$), neutral iron K$\alpha$1 ($6.404 \ \mathrm{keV}$), He-like iron z ($6.637 \ \mathrm{keV}$), He-like iron y ($6.667 \ \mathrm{keV}$), He-like iron x ($6.682 \ \mathrm{keV}$), He-like iron w ($6.700 \ \mathrm{keV}$), H-like iron Ly$\alpha$2 ($6.952 \ \mathrm{keV}$), H-like iron Ly$\alpha$1 ($6.973 \ \mathrm{keV}$), and neutral iron K$\beta$ ($7.058 \ \mathrm{keV}$). These emission lines are broadened due to the Keplerian rotation (\hyperref[Figure0002]{Figure~2c}). Although the intensity of neutral iron K$\alpha$1 is approximately twice that of neutral iron K$\alpha$2, the fluorescence emission lines of neutral iron have a red peak. The origin of this red peak is the neutral iron fluorescence emission line emitted in the outflow. If the angle of inclination is between $50\degr$ and $70\degr$, the solid angle of the outflow on the far side is greater than that of the outflow on the near side.

\subsubsection{Total X-Ray Spectrum}
The observed spectrum is the sum of the transmitted component and the scattered component. \hyperref[Figure0006]{Figure~6} presents the total X-ray spectra for (a) an inclination angle of $i = 55\degr$, (b) $i = 60\degr$, (c) $i = 65\degr$, and (d) $i = 70\degr$. \hyperref[Figure0006]{Figure~6(a)} indicates that the continuum shows absorption lines due to H-like iron and He-like iron, and neutral iron fluorescence emission lines are observed in the case of $i = 55\degr$. If the angle of inclination is between $60\degr$ and $65\degr$, the continuum shows absorption lines due to O-like, N-like, C-like, B-like, Li-like, He-like, and H-like irons (\hyperref[Figure0006]{Figure~6b} and \hyperref[Figure0006]{Figure~6c}). \hyperref[Figure0006]{Figure~6(d)} indicates that only the scattered component is observed due to the presence of Compton-thick material along the line of sight at $i = 70\degr$. In the case of $i \geq 80\degr$, we cannot observe any X-rays because both transmitted and scattered components are absorbed by the Compton-thick material.\clearpage
\begin{figure}\label{Figure0007}
\gridline{\fig{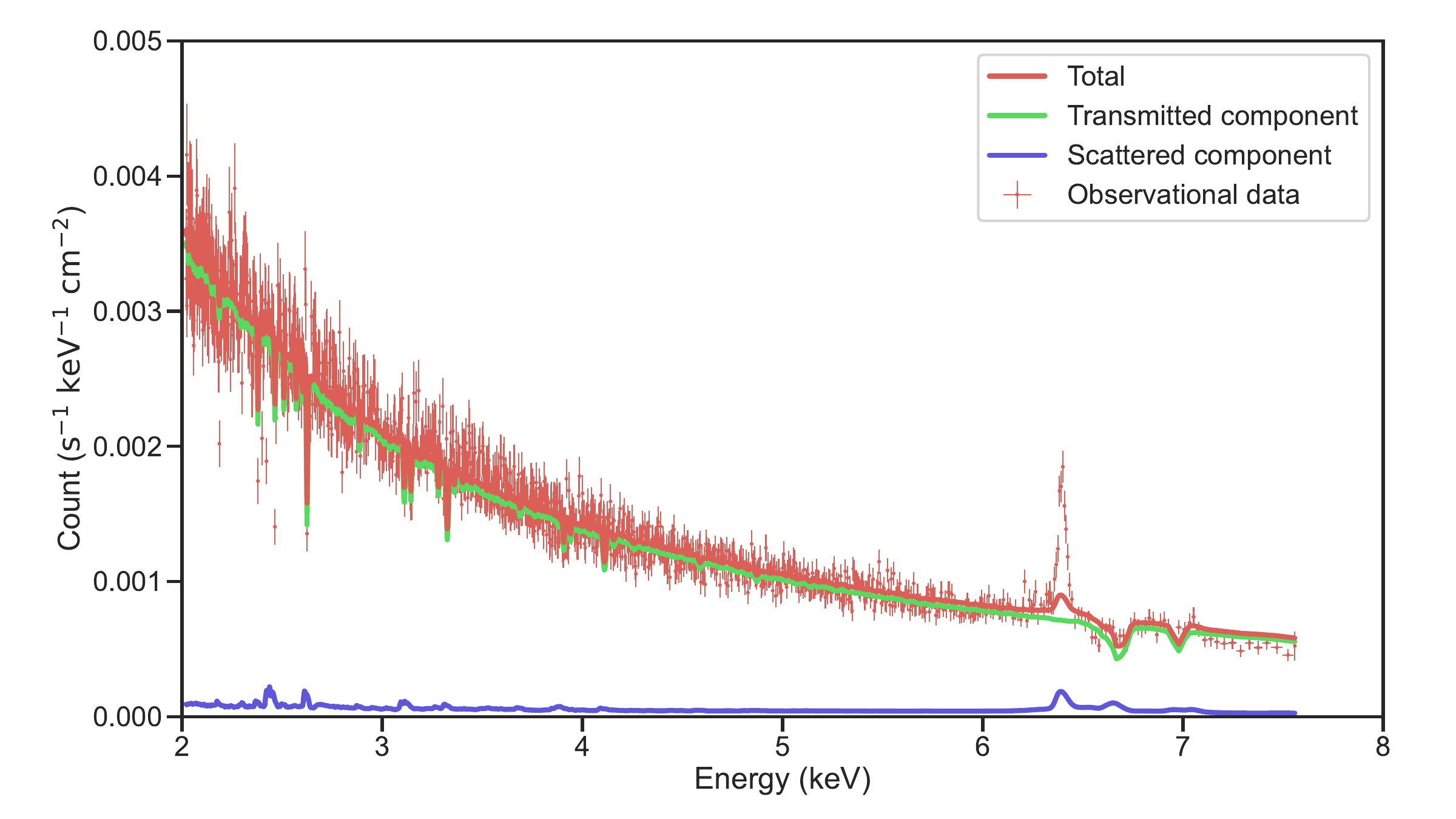}{0.5\textwidth}{(a)}\fig{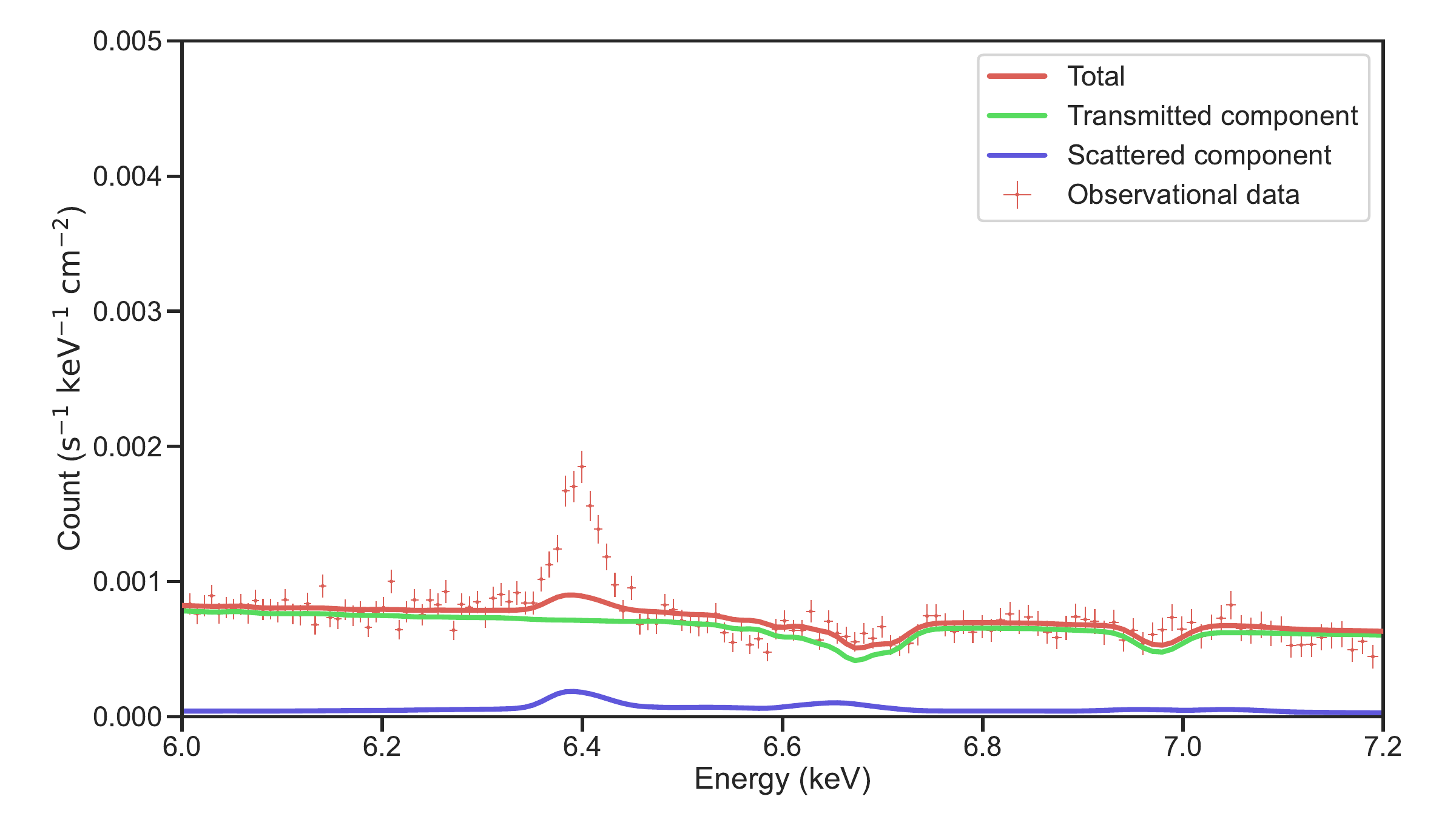}{0.5\textwidth}{(b)}}
\gridline{\fig{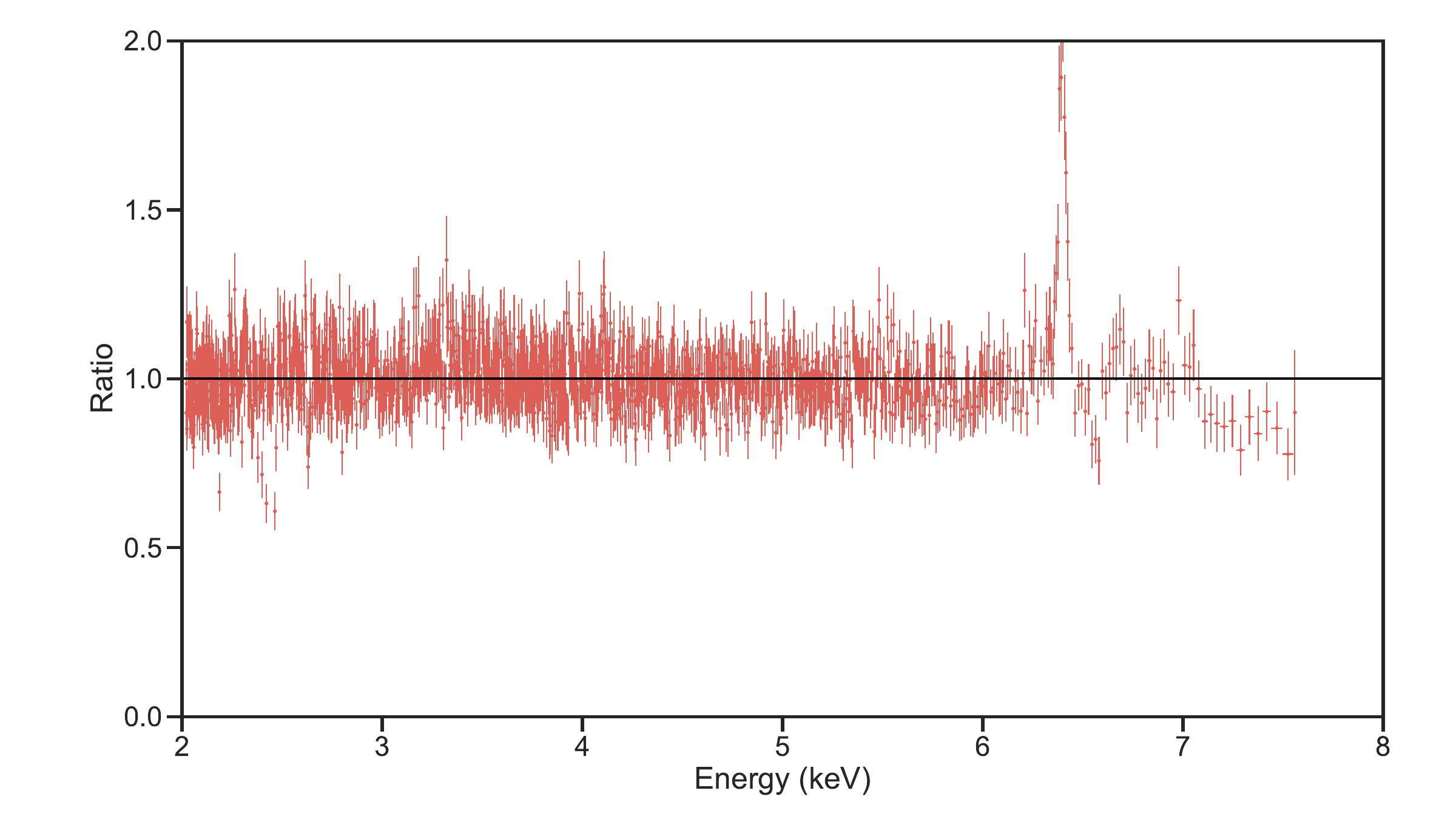}{0.5\textwidth}{(c)}\fig{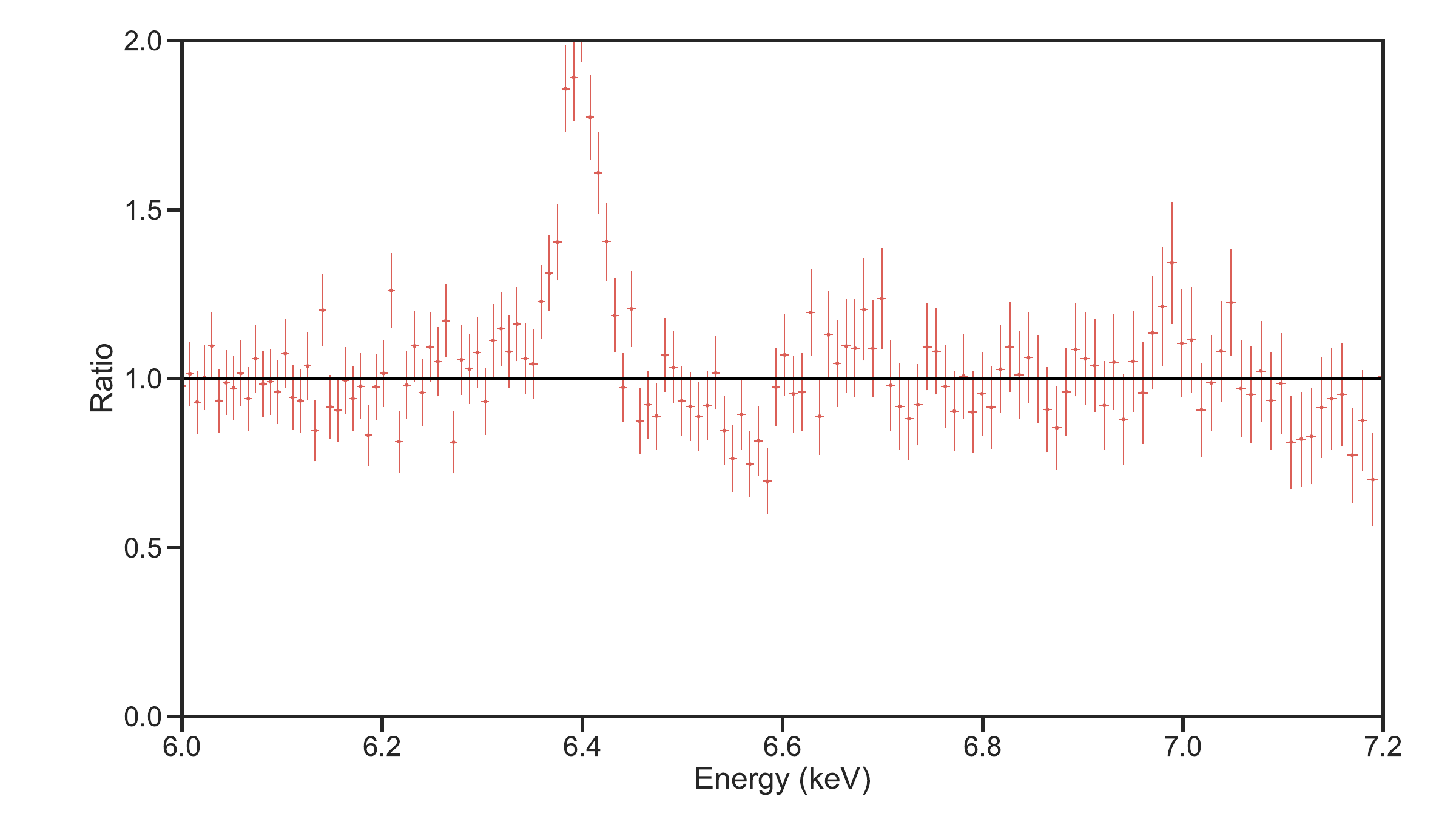}{0.5\textwidth}{(d)}}
\caption{(a) The folded X-ray spectra of NGC~3783 fitted with our model in the rest frame at $2$--$8 \ \mathrm{keV}$. (b) The folded X-ray spectra of NGC~3783 at $6.0$--$7.2 \ \mathrm{keV}$. The red crosses represent the Chandra HETG observational data. The red line shows the total X-ray spectrum. The green line represents the transmitted component obtained from our simulation. The blue line shows the scattered component obtained from our simulation. (c) The ratio of the best-fitting model to the observational data at $2$--$8 \ \mathrm{keV}$. (d) The ratio of the best fitting model to the observational data at $6.0$--$7.2 \ \mathrm{keV}$.}
\end{figure}\clearpage
\subsection{Application to NGC~3783 Observed with Chandra}\label{Section0304}
We applied our model to the X-ray spectrum of NGC~3783 observed with the Chandra High Energy Transmission Grating (HETG; \citealt{Canizares2005}) to investigate whether the X-ray spectral model derived from our simulations can explain the observed X-ray spectrum. We selected NGC~3783 because its SMBH mass and the Eddington ratio are nearly identical to those assumed in our radiative hydrodynamic simulations. NGC~3783 is a luminous AGN located in the nearby universe, at a distance of $39.9_{-11.9}^{+14.5} \ \mathrm{Mpc}$ \citep{Gravity2021}. NGC~3783 hosts an SMBH with a mass of $2.54_{-0.72}^{+0.90} \times 10^{7} M_{\sun}$ \citep{Gravity2021} and an Eddington ratio of approximately $10\%$ \citep{Koss2017}.

\subsubsection{Data Analysis}
The Chandra observed NGC~3783 six times between 2000 and 2001. \cite{Kaspi2002} indicated that the absorption lines are blue shifted relative to the systemic velocity by a mean velocity of $590 \pm 150 \ \mathrm{km}/\mathrm{s}$. Since NGC~3783 has been observed by Chandra HETG for $900 \ \mathrm{ksec}$, X-ray spectra based on various theories have been compared with X-ray spectroscopic observations by Chandra HETG \citep[e.g.,][]{Kaspi2000, Kaspi2001, Kaspi2002, Krongold2003, Krongold2005, Netzer2003, Chelouche2005, Goosmann2016, Mehdipour2017, Mao2019}. We obtained these observational data from the Chandra Transmission Grating Data Archive and Catalog (TGCat; \citealt{Huenemoerder2011}). To improve photon statistics, we combined these six observations, totaling $900 \ \mathrm{ksec}$. In this analysis, we used Chandra Interactive Analysis of Observations version 4.16 and the calibration database version 4.11.2 \citep{Fruscione2006}.

\subsubsection{Spectral Analysis}
We applied the X-ray spectral model in the XSPEC terminology as follows.
\begin{equation}
\mathrm{phabs}*(\mathrm{atable\{fountain\_transmitted.fits\}} + \mathrm{atable\{fountain\_scattered.fits\}})
\end{equation}
The phabs represents the Galactic absorption. We adopted a hydrogen column density of $N_{\mathrm{H}} = 1.01 \times 10^{21} \ \mathrm{cm}^{-2}$ based on \cite{HI4PI2016}. The $\mathrm{atable\{fountain\_transmitted.fits\}}$ and $\mathrm{atable\{fountain\_scattered.fits\}}$ denote the transmitted component and the scattered component derived from the simulation, respectively. These models include four parameters: (1) the photon index $\Gamma$, (2) the angle of inclination $i$, (3) the redshift, and (4) normalization. We adopted a redshift $z = 0.009730$ based on the NASA/IPAC Extragalactic Database. The photon index, angle of inclination, and normalization are treated as free parameters.

\subsubsection{Comparison between Theoretical Model and Observational Data}
We applied the model to the $2.0$--$7.5 \ \mathrm{keV}$ X-ray spectra observed with Chandra HETG. Since the range of the angle of inclination in which the absorption lines are observed is from $55\degr$ to $65\degr$, we used the stepper command to move the angle of inclination from $55\degr$ to $65\degr$ by $0.1\degr$. \hyperref[Figure0007]{Figure~7(a)} shows the folded X-ray spectra of NGC~3783 fitted with the model in the rest frame at $2$--$8 \ \mathrm{keV}$. The X-ray spectra of NGC~3783 exhibits the absorption lines due to silicon, sulfur, argon, calcium, iron, and nickel. \hyperref[Figure0007]{Figure~7(b)} shows an enlarged $6.0$--$7.2 \ \mathrm{keV}$ X-ray spectra. \hyperref[Figure0007]{Figure~7(b)} indicates that the X-ray spectra of NGC~3783 shows emission lines of neutral iron fluorescence and absorption lines due to He-like and H-like irons. We note that since the energy resolution of the Chandra HETG does not resolve absorption lines due to Be-like, Li-like, or He-like irons, these absorption lines may be mixed in.

\hyperref[Figure0007]{Figure~7(c)} indicates that our model is capable of reproducing the X-ray spectra of NGC~3783, attaining a fit statistic of $C/\mathrm{dof} = 2491/1817$. The best fitting parameters are the photon index of $\Gamma = 1.45_{-0.01}^{+0.01}$, the angle of inclination $i$ of $56.8\degr_{-0.2}^{+0.2}$ and the normalization of $1.28_{-0.01}^{+0.01} \times 10^{-2} \ \mathrm{photons} \ \mathrm{keV}^{-1} \ \mathrm{cm}^{-2} \ \mathrm{s}^{-1}$. We note that the error on a parameter corresponds to a $90\%$ credible interval, and we estimate the error using a Markov chain Monte Carlo with a length of 1,000,000. Our model roughly reproduces absorption lines due to He-like and H-like irons (\hyperref[Figure0007]{Figure~7d}). If the angle of inclination is approximately $55\degr$, the line of sight outflow velocity is approximately $500 \ \mathrm{km} \ \mathrm{s}^{-1}$, which is comparable to the velocity of the highly ionized component of $480\pm10 \ \mathrm{km} \ \mathrm{s}^{-1}$ \citep{Mao2019}. However, our model underestimates the flux of neutral iron fluorescent emission lines. This is because our computational domain only considers areas inside $0.02 \ \mathrm{pc}$. The presence of spatially extended neutral iron fluorescence emission lines has been proposed by high spatial resolution observations of Chandra \citep{Fabbiano2017, Kawamuro2019, Kawamuro2020}. This suggests that a radiative hydrodynamic simulation that considers a wider region is needed to explain the flux of neutral iron fluorescence emission lines.  
\begin{figure}\label{Figure0008}
\gridline{\fig{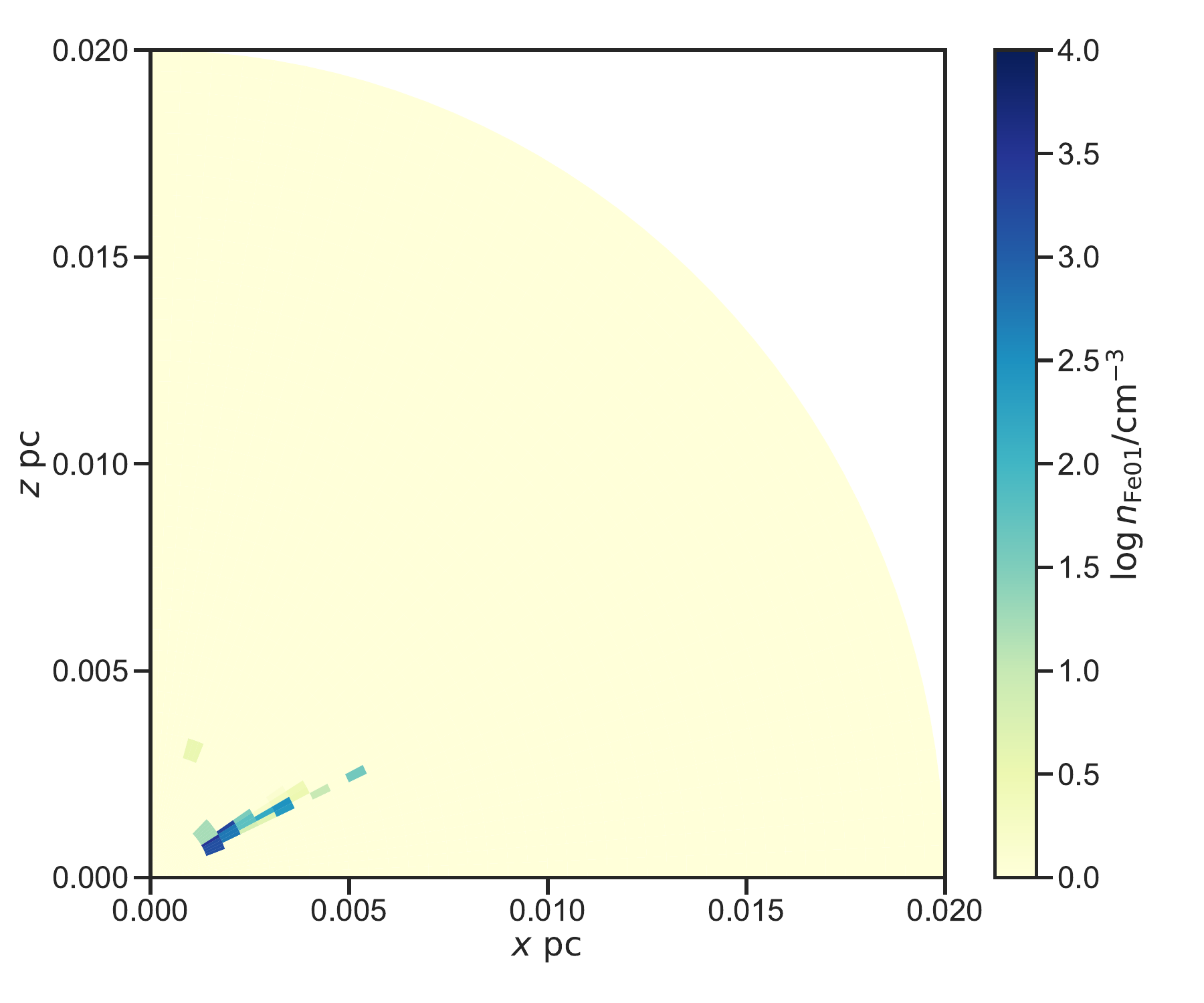}{0.5\textwidth}{(a)}}
\caption{The distribution of the number density of H-like iron $\log n_{\mathrm{Fe01}}/\mathrm{cm}^{-3}$.}
\end{figure}
\begin{figure}\label{Figure0009}
\gridline{\fig{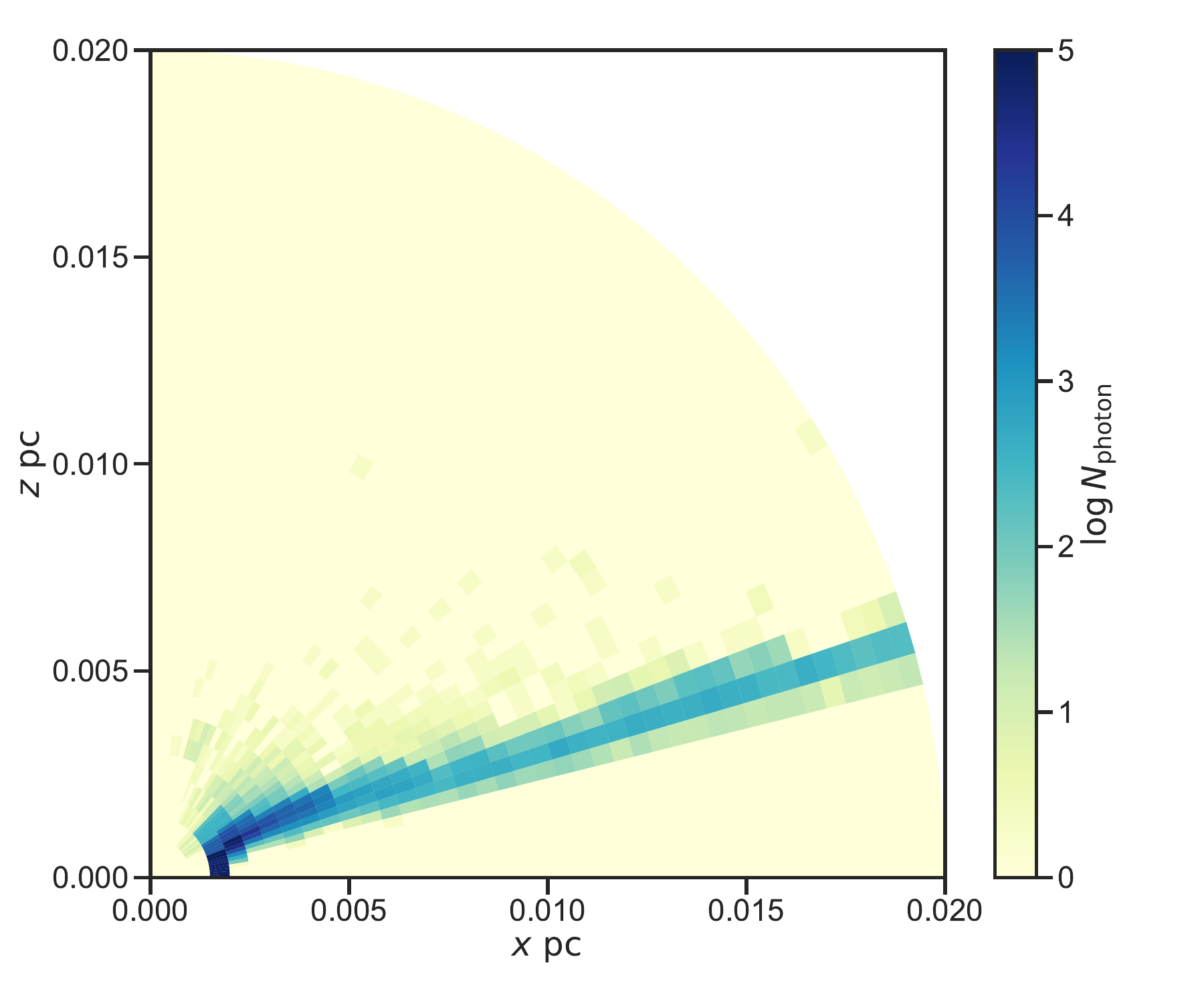}{0.5\textwidth}{}}
\caption{The distribution of neutral iron fluorescent emission lines.}
\end{figure}
\begin{figure}\label{Figure0010}
\gridline{\fig{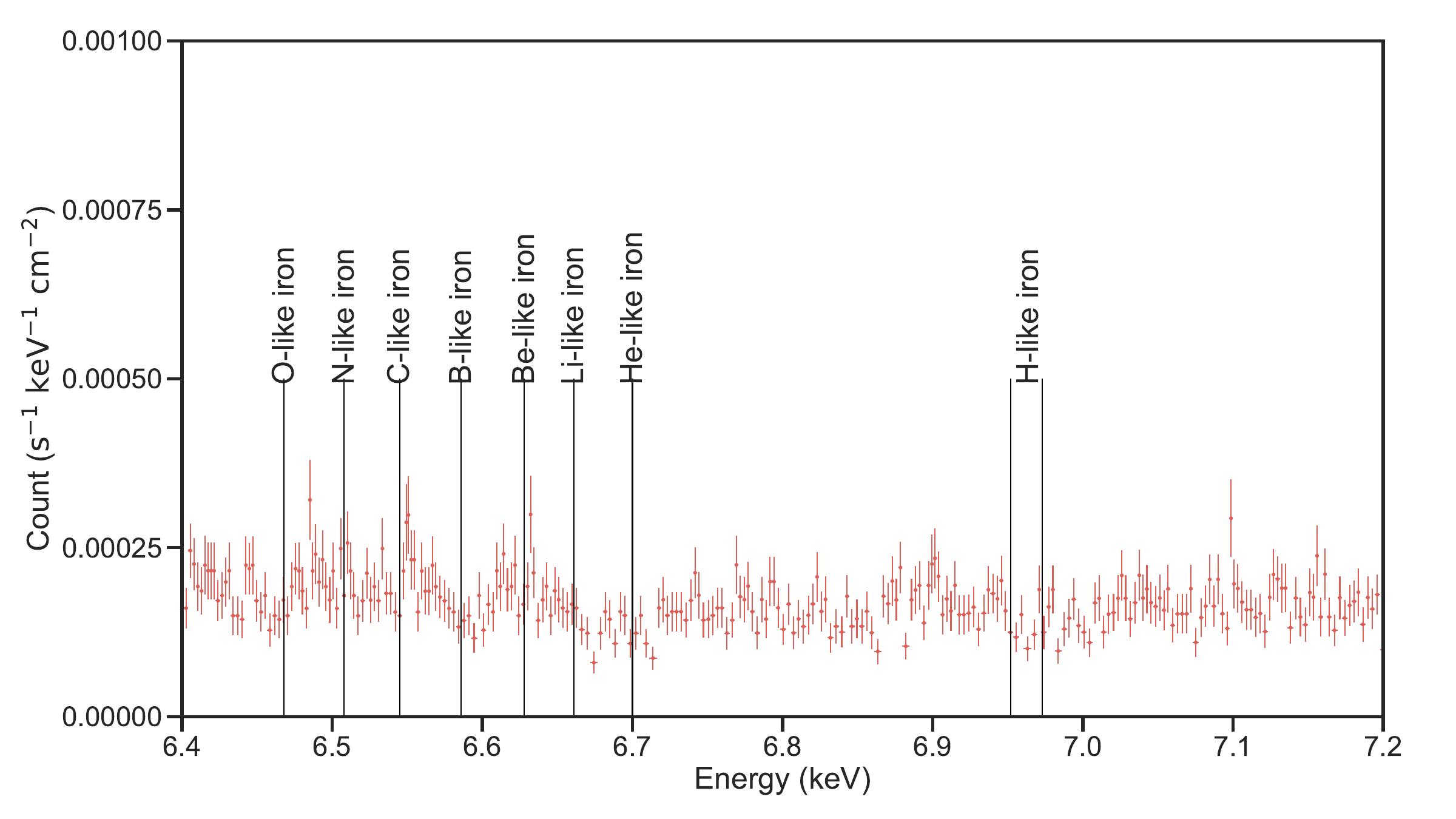}{0.5\textwidth}{}}
\caption{The simulation results for $320 \ \mathrm{ksec}$ observation by XRISM.}
\end{figure}\clearpage
\section{Discussion}\label{Section0400}
We performed the X-ray radiative transfer based on the three-dimensional radiative hydrodynamic simulations on the sub-parsec scale and applied our model to the X-ray spectrum of NGC~3783 observed with the Chandra HETG. In this section, we discuss the origin of H-like iron absorption lines (\hyperref[Section0401]{Section~4.1}), the origin of neutral iron fluorescence emission lines (\hyperref[Section0402]{Section~4.2}), and the observational simulations with the X-Ray Imaging and Spectroscopy Mission (XRISM) (\hyperref[Section0403]{Section~4.3}).
\subsection{Origin of H-like Iron Absorption Lines}\label{Section0401}
To determine where H-like iron absorption lines are produced, we examined the number density distribution of H-like iron based on the results of the photoionization equilibrium calculation. \hyperref[Figure0008]{Figure~8} presents the distribution of the number density of H-like iron $\log n_{\mathrm{Fe01}}/\mathrm{cm}^{-3}$. Here we assumed the solar abundance of \cite{Anders1989} to calculate the number density of H-like iron from the hydrogen number density. \hyperref[Figure0008]{Figure~8} indicates that H-like iron exists within a radius of $0.005 \ \mathrm{pc}$ and an angle of inclination of $55\degr$ to $65\degr$. We note that it is difficult to discuss the fraction of sources for which absorption lines are observed from this calculation alone. Although the Eddington ratio is assumed to be $10\%$ in this calculation, if this value is changed, the angle of inclination range in which the absorption lines are observed also changes. To estimate the fraction from theoretical calculations, radiative hydrodynamic simulations for various Eddington ratios are essential, which is beyond the scope of this paper.
\subsection{Origin of Iron Fluorescence Lines}\label{Section0402}
The neutral iron fluorescent emission lines are an important clue to the structure of AGN. The X-Ray Imaging and Spectroscopy Mission (XRISM; \citealt{Tashiro2025}), launched in 2023, is equipped with a microcalorimeter detector (Resolve; \citealt{Ishisaki2022}). Since the energy resolution of Resolve at $6 \ \mathrm{keV}$ is approximately $30$ times higher than that of the CCD detectors, precise X-ray spectroscopy with Resolve is one of the best methods to measure the profile of the neutral iron fluorescence emission lines. XRISM observed Seyfert~1 galaxy NGC~4151 in December 2023 and showed that the profile consists of at least two components, a narrow component of less than $300 \ \mathrm{km} \ \mathrm{s}^{-1}$ and a broad component of approximately $5000 \ \mathrm{km} \ \mathrm{s}^{-1}$ \citep{XRISM2024}.

Since MONACO has information on where the neutral iron fluorescent emission lines were emitted, we investigate the origin of the neutral iron fluorescent emission lines based on three-dimensional radiative hydrodynamic simulations. \hyperref[Figure0009]{Figure~9} presents the distribution of neutral iron fluorescent emission lines when observed from an inclination angle $i$ from $30\degr$ to $60\degr$. \hyperref[Figure0009]{Figure~9} indicates that neutral iron fluorescence emission lines are emitted from the surface of Compton-thick materials, especially from regions with radii smaller than $0.005 \ \mathrm{pc}$. In fact, the Keplerian rotation velocity in the $0.005 \ \mathrm{pc}$ radius region is approximately $7000 \ \mathrm{km} \ \mathrm{s}^{-1}$. This component corresponds to the broad component observed in NGC~4151.
\subsection{XRISM Simulation}\label{Section0403}
Since XRISM observed NGC~3783 for $320 \ \mathrm{ksec}$ in 2024 July, we finally simulated the XRISM observation using the best-fitting parameters obtained from the spectral analysis. \hyperref[Figure0010]{Figure~10} shows the simulation results for the $320 \ \mathrm{ksec}$ observation by the XRISM satellite. Although absorption lines from Be-like iron, Li-like iron, and He-like iron could not be resolved in the case of Chandra HETG, XRISM Resolve may be able to resolve these absorption lines and the profile of the neutral iron fluorescence emission lines. In other words, XRISM Resolve observations allow us to study the ionization state of the outflow and the driving mechanism of the outflow in more detail.\clearpage
\section{Conclusion}\label{Section0500}
Although the X-ray spectra of Seyfert~1 galaxies exhibit absorption lines due to He-like and H-like irons at blue-shifted velocities of approximately $500 \ \mathrm{km} \ \mathrm{s}^{-1}$, the physical origin of these absorption lines remains uncertain. In this study, we performed X-ray radiative transfer calculations based on the sub-parsec-scale three-dimensional radiative hydrodynamic simulations. The key findings are as follows.

\begin{enumerate}
\item The transmitted component of the X-ray spectrum varies with both the angle of inclination and the azimuthal angle. When the angle of inclination ranges from $55\degr$ to $65\degr$, the transmitted component exhibits the absorption lines of O-like, N-like, C-like, B-like, Li-like, He-like, and H-like irons. These absorption lines show blue shifts of approximately $500 \ \mathrm{km} \ \mathrm{s}^{-1}$ and originate from regions within $0.005 \ \mathrm{pc}$.

\item The scattered component of the X-ray spectrum exhibits emission lines of neutral iron K$\alpha$, He-like iron (z, y, x, w), H-like iron Ly$\alpha$, and neutral iron K$\beta$. These emission lines are broadened by the Keplerian rotation of approximately $7000 \ \mathrm{km} \ \mathrm{s}^{-1}$ and are emitted from regions within $0.005 \ \mathrm{pc}$.

\item We applied our model to the X-ray spectrum of NGC~3783 observed by the Chandra HETG. Our model reproduced the continuum and absorption lines of He-like and H-like irons.

\item We simulated the XRISM observation based on the best-fitting parameters obtained from the spectral analysis. Although the absorption lines of Be-like iron, Li-like iron, and He-like iron could not be resolved in the case of Chandra HETG, XRISM Resolve may be able to resolve these absorption lines.
\end{enumerate}
\begin{acknowledgements}
We thank the anonymous referee who provided useful and detailed comments. Atsushi Tanimoto and the present research are partly supported by the Kagoshima University postdoctoral research program (KU-DREAM). This work is also supported by the Grant-in-Aid for Early Career Scientists Grant No. 23K13147 (AT) and the Grants-in-Aid for Scientific Research Grant Nos. 21H04496 (KW), 22H00128 (HO), 24K17080 (YK), and 19K03918 (NK). Numerical computations were performed on Cray XC50 at the Center for Computational Astrophysics, National Astronomical Observatory of Japan. This research has also used the NASA/IPAC extragalactic database, which is operated by the Jet Propulsion Laboratory, California Institute of Technology, under contract with the National Aeronautics and Space Administration.
\end{acknowledgements}
\facilities{Chandra}
\software{HEASoft 6.34, MONACO \citep{Odaka2016}, XSPEC \citep{Arnaud1996}}\clearpage
\bibliography{tanimoto}
\bibliographystyle{aasjournal}
\end{document}